\documentclass{llncs}

\usepackage[all]{xy}
\usepackage[mathscr]{euscript}
\usepackage{graphics}
\usepackage{graphicx}
\usepackage{ae}
\usepackage{aecompl}
\usepackage{pslatex}
\usepackage{xspace}
\usepackage{amsmath,amsfonts,amstext,amssymb,verbatim}
\usepackage[english]{babel}
\usepackage[T1]{fontenc}
\usepackage{float}
\usepackage{latexsym}
\usepackage{url}
\usepackage{aecompl}
\usepackage{wasysym}
\usepackage{times}

\newcommand{\GAP}{0.5em}
\newcommand{\one}{\ofont{1}}
\newcommand{\zero}{\ofont{0}}
\newcommand{\mysubsection}[1]{\vspace{\GAP}\noindent\textbf{#1}\,}
\newcommand{\myparagraph}[1]{\vspace{\GAP}\noindent\textit{#1}\,}

\newif\iflncs\lncstrue
\newif\ifmai\maifalse

\newcommand{\act}{\mu}
\newcommand{\AtCCS}{\textsc{AtCCS}\xspace}

\newcommand{\Alt}{\;\;\mbox{\rule[-.35ex]{.2ex}{1em}}\;\;}
\newcommand*{\parall}{\mathbin{\,{\mid}\,}}
\newcommand{\rulen}[1]{\ensuremath{\textsc{#1}}}
\newcommand{\pv}{\,{\mathtt{;}}\,}
\newcommand{\Res}[2]{\setminus^{{#1}}{#2}}
\newcommand{\Si}{\sigma} \newcommand{\T}{\theta}
\newcommand{\D}{\delta} \newcommand{\nil}{\mathbf{0}}
\newcommand{\ofont}[1]{\mathtt{#1}}
\newcommand{\End}{\ofont{end}}
\newcommand{\Retry}{\ofont{retry}}

\newcommand{\Orelse}{\;\ofont{orElse}\;}
\newcommand{\Oorelse}{\Orelse}
\newcommand{\Read}[1]{\ofont{rd}\:{#1}\,}
\newcommand{\Write}[1]{\ofont{wt}\:{#1}\,}
\newcommand{\ReadL}[1]{\text{\textsc{rd}}{#1}\,}
\newcommand{\WriteL}[1]{\text{\textsc{wt}}\:{#1}\,}
\newcommand{\Inp}[1]{\mathbin{#1}} 
\newcommand{\Out}[1]{\mathbin{\overline{#1}}} 
\newcommand{\Atomic}{\ofont{atom}} \newcommand{\Eblock}[2]{
  ({#1})_{#2}} \newcommand{\Ablock}[2]{\{\!\!|{#1} |\!\!\}_{{#2}}}
\newcommand{\tob}{\preceq_{0}}
\newcommand{\tot}{\preceq}
\newcommand{\mayeq}{\smash[b]{\simeq_{may}}} 
\newcommand{\maypr}{\smash[b]{\:\raisebox{-.7ex} 
{$\,{\stackrel{\displaystyle \sqsubset}{\sim}}_{may} \,$}\:}}
\newcommand{\bisim}{\approx}
\newcommand{\nbisim}{\not\approx}
\newcommand{\Abisim}{\approx_a} 
\newcommand{\nAbisim}{{\not\approx}_a} 
\newcommand{\AAmore}{\sqsupseteq}
\newcommand{\AAbisim}{\backsimeq} 
\newcommand{\nAAbisim}{\not\backsimeq} 
\newcommand{\Ll}{\ll_{may}} 
\newcommand{\tradm}[1]{[\![ {#1} ]\!]}

\makeatletter
\def \rightarrowfill{\m@th\mathord{\smash-}\mkern-6mu%
  \cleaders\hbox{$\mkern-2mu\mathord{\smash-}\mkern-2mu$}\hfill
  \mkern-6mu\mathord\rightarrow}
\makeatother

\newcommand{\tom}{\to} 
\newcommand{\tolabel}[1]{\smash[t]{\overstackrel{\rightarrowfill}{\ #1\ }}} 
\newcommand{\Tolabel}[1]{\smash[t]{\overstackrel{\Rightarrowfill}{\ #1\ }}}
\newcommand{\To}{\Rightarrow}

\iflncs{}\else \newtheorem{definition}{Definition}\fi
\iflncs{}\else \newtheorem{proposition}{Proposition}\fi
\newtheorem{proposition_a}{Proposition}[section]
\iflncs{}\else \newtheorem{theorem}{Theorem}\fi
\newtheorem{theorem_a}{Theorem}[section]
\iflncs{}\else  \fi
\newtheorem{corollary_a}{Corollary}[section]
\iflncs{}\else \newtheorem{lemma}{Lemma}\fi
\newtheorem{lemma_a}{Lemma}[section]
\iflncs{}\else \fi
\iflncs{}\else \fi
\newcommand{\defi}{\stackrel{\smash[b]{\vartriangle}}{=}}
\newcommand{\overstackrel}[2]{\mathrel{\mathop{#1}\limits^{#2}}}
\newcommand{\observer}{\mathcal{O}}
\newcommand{\Obs}{\ensuremath{\mathcal{O}\mathit{bs}}\xspace}
\newcommand{\System}[5]{D\{{#1};{#2};{#3};{#4};{#5}\}}

\makeatletter
\def \rightarrowfill{\m@th\mathord{\smash-}\mkern-6mu%
  \cleaders\hbox{$\mkern-2mu\mathord{\smash-}\mkern-2mu$}\hfill
  \mkern-6mu\mathord\rightarrow}
\makeatother

\makeatletter
\def \Rightarrowfill{\m@th\mathord{\smash-}\mkern-6mu%
  \cleaders\hbox{$\mkern-2mu\mathord{\smash-}\mkern-2mu$}\hfill
  \mkern-6mu\mathord\Rightarrow}
\makeatother

\newcommand{\deduce}[2]{\frac{\displaystyle #1}{\displaystyle #2}}

\makeatletter
\def \rightarrowfill{\m@th\mathord{\smash-}\mkern-6mu%
  \cleaders\hbox{$\mkern-2mu\mathord{\smash-}\mkern-2mu$}\hfill
  \mkern-6mu\mathord\rightarrow}
\makeatother

\makeatletter
\def \Rightarrowfill{\m@th\mathord{\smash=}\mkern-6mu%
  \cleaders\hbox{$\mkern-2mu\mathord{\smash=}\mkern-2mu$}\hfill
  \mkern-6mu\mathord\Rightarrow}
\makeatother

\makeatletter
\def \midrightarrowfill{\m@th\mathord{\smash{\raisebox{.2ex}{$\scriptscriptstyle\mid$}}\!\!\,-}\mkern-6mu%
  \cleaders\hbox{$\mkern-2mu\mathord{\smash-}\mkern-2mu$}\hfill
  \mkern-6mu\mathord\rightarrow}
\makeatother

\makeatletter
\def \midRightarrowfill{\m@th\mathord{\smash{\raisebox{.1ex}{$\scriptstyle\mid$}}\!\!\!=}\mkern-6mu%
  \cleaders\hbox{$\mkern-2mu\mathord{\smash=}\mkern-2mu$}\hfill
  \mkern-6mu\mathord\Rightarrow}
\makeatother

\begin{document}
\title{A Concurrent Calculus with Atomic Transactions}
\author{Lucia Acciai\inst{1} 
  \and Michele Boreale\inst{2} 
  \and Silvano Dal Zilio\inst{1}}
\institute{Laboratoire d'Informatique Fondamentale de Marseille (LIF),\\
  CNRS and Universit\'e de Provence, France \and Dipartimento di
  Sistemi e Informatica, Universit\`a di Firenze, Italy}
\maketitle
\setcounter{footnote}{0}
\begin{abstract}
  The \emph{Software Transactional Memory} (STM) model is an original
  approach for controlling concurrent accesses to ressources without
  the need for explicit lock-based synchronization mechanisms. A key
  feature of STM is to provide a way to group sequences of read and
  write actions inside \emph{atomic blocks}, similar to database
  transactions, whose whole effect should occur atomically.

  In this paper, we investigate STM from a process algebra perspective
  and define an extension of asynchronous CCS with atomic blocks of
  actions. Our goal is not only to set a formal ground for reasoning
  on STM implementations but also to understand how this model fits
  with other concurrency control mechanisms. We also view this
  calculus as a test bed for extending process calculi with atomic
  transactions. This is an interesting direction for investigation
  since, for the most part, actual works that mix transactions with
  process calculi consider compensating transactions, a model that
  lacks all the well-known ACID properties.

  We show that the addition of atomic transactions results in a very
  expressive calculus, enough to easily encode other concurrent
  primitives such as guarded choice and multiset-synchronization (à la
  join-calculus). The correctness of our encodings is proved using a
  suitable notion of bisimulation equivalence. The equivalence is then
  applied to prove interesting ``laws of transactions'' and to obtain
  a simple normal form for transactions.
\end{abstract}
\section{Introduction}
The craft of programming concurrent applications is about mastering
the strains between two key factors: getting hold of results as
quickly as possible, while ensuring that only correct results (and
behaviors) are observed. To this end, it is vital to avoid unwarranted
access to shared resources. The \emph{Software Transactional Memory}
(STM) model is an original approach for controlling concurrent
accesses to resources without using explicit lock-based
synchronization mechanisms. Similarly to database transactions, the
STM approach provides a way to group sequences of read and write
actions inside \emph{atomic blocks} whose whole effect should occur
atomically. The STM model has several advantages. Most notably, it
dispenses the programmer with the need to explicitly manipulate locks,
a task widely recognized as difficult and error-prone. Moreover,
atomic transactions provide a clean conceptual basis for concurrency
control, which should ease the verification of concurrent programs.
Finally, the model is effective: there exist several STM
implementations for designing software for multiprocessor systems;
these applications exhibit good performances in practice (compared to
equivalent, hand-crafted, code using locks).

We investigate the STM model from a process algebra perspective and
define an extension of asynchronous CCS~\cite{accs} with atomic blocks
of actions. We call this calculus \AtCCS. The choice of a dialect of
CCS is motivated by an attention to economy: to focus on STM
primitives, we study a calculus as simple as possible and dispense
with orthogonal issues such as values, mobility of names or processes,
\emph{etc}. We believe that our work could be easily transferred to a
richer setting. Our goal is not only to set a formal ground for
reasoning on STM implementations but also to understand how this model
fits with other concurrency control mechanisms. We also view this
calculus as a test bed for extending process calculi with atomic
transactions. This is an interesting direction for investigation
since, for the most part, works that mix transactions with process
calculi consider \emph{compensating transactions}, see
e.g.~\cite{BCGHM,BPG05,BLZ03,BMM04,BMM05,ccsp,BHF04,BFN05,webpi}.

The idea of providing hardware support for software transactions
originated from works by Herlihy and Moss~\cite{STM0} and was later
extended by Shavit and Touitou~\cite{STMST} to software-only
transactional memory. Transactions are used to protect the execution
of an atomic block. Intuitively, each thread that enters a transaction
takes a snapshot of the shared memory (the global state). The
evaluation is optimistic and all actions are performed on a copy of
the memory (the local state). When the transaction ends, the snapshot
is compared with the current state of the memory. There are two
possible outcomes: if the check indicates that concurrent writes have
occurred, the transaction aborts and is rescheduled; otherwise, the
transaction is committed and its effects are propagated
instantaneously. Very recently, Harris et al.~\cite{STM} have proposed
a (combinator style) language of transactions that enables arbitrary
atomic operations to be composed into larger \emph{atomic
  expressions}. We base the syntax of \AtCCS on the operators defined
in~\cite{STM}.

The main contributions of this work are: (1) the definition of a
process calculus with atomic transactions; and (2) the definition of
an asynchronous bisimulation equivalence $\Abisim$ that allows
compositional reasoning on transactions. We also have a number of more
specific technical results. We show that \AtCCS is expressive enough
to easily encode interesting concurrent primitives, such as
(preemptive versions of) guarded choice and multiset-synchronization,
and the leader election problem (Section~\ref{sec:encod-conc-prim}).
Next, we define an equivalence between atomic expressions $\AAbisim$
and prove that $\Abisim$ and $\AAbisim$ are congruences
(Section~\ref{sec:bisimulation}).  These equivalences are used to
prove the correctness of our encodings, to prove interesting
``behavioral laws of transactions'' and to define a simple normal form
for transactions. We also show that transactions (modulo $\AAbisim$)
have an algebraic structure close to that of a {bound semilattice}, an
observation that could help improve the design of the transaction
language. Finally, we propose a may-testing equivalence for \AtCCS,
give an equivalent characterization using a trace-based semantics and
show that may testing equivalence is unable to notice the presence of
transactions (Section~\ref{sec:trace}).  Section~\ref{sec:conclusions}
concludes with an overview on future and related works. The proofs of the main results are reported in the appendices.

\section{The calculus}\label{sec:basiccalculus}
We define the syntax and operational semantics of \AtCCS, which is
essentially a cut down version of asynchronous CCS, without choice and
relabeling operators, equipped with atomic blocks and constructs for
composing (transactional) sequences of actions.

\mysubsection{Syntax of Processes and Atomic Expressions.}
The syntax of \AtCCS, given in Table~\ref{Syntax}, is divided into
syntactical categories that define a stratification of terms.  The
definition of the calculus depends on a set of names, ranged over by
$a,\,b,\,\dots$ As in CCS, names model communication channels used in
process synchronization, but they also occur as objects of read and
write actions in atomic transactions.

\emph{Atomic expressions}, ranged over by $M,\,N,\, \dots$, are used
to define sequences of actions whose effect should happen atomically.
Actions $\Read a$ and $\Write a$ represent attempts to input and
output to the channel $a$. Instead of using snapshots of the state for
managing transaction, we use a log-based approach. During the
evaluation of an atomic block, actions are recorded in a private log
$\D$ (a sequence $\alpha_1 \dots \alpha_n$) and have no effects
outside the scope of the transaction until it is committed.  The
action $\Retry$ aborts an atomic expression unconditionally and starts
its execution afresh, with an empty log $\epsilon$. The termination
action $\End$ signals that an expression is finished and should be
committed. If the transaction can be committed, all actions in the log
are performed at the same time and the transaction is closed,
otherwise the transaction aborts. Finally, transactions can be
composed using the operator $\ofont{orElse}$, which implements
(preemptive) alternatives between expressions. $M \Orelse N$ behaves as
expression $N$  if $M$ aborts and has the behavior of $M$
otherwise.

\emph{Processes}, ranged over by $P,\,Q,\, R,\, \dots$, model
concurrent systems of communicating agents. We have the usual
operators of CCS: the empty process, $\nil$, the parallel composition
$P\parall Q$, and the input prefix $\Inp a.P$. There are some
differences though. The calculus is asynchronous, meaning that a
process cannot block on output actions.  Also, we use \emph{replicated
  input} $*\Inp a .P$ instead of recursion (this does not change the
expressiveness of the calculus) and we lack the choice and relabeling
operators of CCS.  Finally, the main addition is the presence of the
operator $\Atomic(M)$, which models a transaction that safeguards the
expression $M$. The process $\Ablock{A}{M}$ represents the ongoing
evaluation of an atomic block $M$: the subscript is used to keep the
initial code of the transaction, in case it is aborted and executed
afresh, while $A$ holds the remaining actions that should be
performed.

\begin{table}[!t]
  \centering{\small
    \framebox{$\begin{array}{rrrll}
        \text{Actions} & \alpha,\beta & ::= & \Read  a &
        \text{(tentative) read access to $a$}\\
        && \Alt & \Write a   & \text{(tentative) write access to $a$}      \\[\GAP]
        \textrm{(Atomic) Expressions} &M,N & ::= & \End  & \text{termination}\\
        && \Alt & \Retry   & \text{abort and retry the current atomic block}\\
        && \Alt & \alpha.M  & \text{action prefix}     \\
        && \Alt & M \Orelse N  \quad & \text{alternative}   \\[\GAP]
        \textrm{Ongoing expressions} 
        &A,B & ::= & \Eblock{M}{\Si;\D}  & \text{execution of $M$ with state
          $\sigma$ and log $\delta$}      \\
        && \Alt &  A\Oorelse B & \text{ongoing alternative}\\[\GAP]
        \textrm{Processes} &P,Q & ::= & \nil  & \text{nil}     \\
        & & \Alt & \Out a  & \text{(asynchronous) output} \\
        & & \Alt & \Inp a .P & \text{input} \\
        & & \Alt & *\Inp a .P & \text{replicated input} \\
        & & \Alt & P \parall Q & \text{parallel composition} \\
        & & \Alt & P\Res{n} a  & \text{hiding} \\
        & & \Alt & \Atomic (M) & \text{atomic block}  \\
        & & \Alt & \Ablock {A}{M} & \text{ongoing atomic block} \\
        \end{array}$ }\vspace*{\GAP}} 
  \caption{Syntax of \protect{\AtCCS}: Processes and Atomic Expressions.}\label{Syntax}
\end{table}

An \emph{ongoing atomic block}, $A, B, \dots$, is essentially an
atomic expression enriched with an \emph{evaluation state} $\Si$ and a
\emph{log} $\D$ of the currently recorded actions. A state $\Si$ is a
multiset of names that represents the output actions visible to the
transaction when it was initiated. (This notion of state bears some
resemblance with tuples space in coordination calculi, such as
Linda~\cite{linda}.) When a transaction ends, the state $\Si$ recorded
in the block $\Eblock{M}{\Si;\D}$ (the state at the initiation of the
transaction) can be compared with the current state (the state when
the transaction ends) to check if other processes have concurrently
made changes to the global state, in which case the transaction should
be aborted.

\mysubsection{Notation.} In the following, we write $\Si \uplus \{ a
\}$ for the multiset $\Si$ enriched with the name $a$ and $\Si
\setminus \Si'$ for the multiset obtained from $\Si$ by removing
elements found in $\Si'$, that is the smallest multiset $\Si''$ such
that $\Si \subseteq \Si' \uplus \Si''$.  The symbol $\emptyset$ stands
for the empty multiset while $\{ a^n \}$ is the multiset composed of
exactly $n$ copies of $a$, where $\{ a^0 \} = \emptyset$.

Given a log $\D$, we use the notation $\WriteL (\D)$ for the multiset
of names which appear as objects of a write action in $\D$.
Similarly, we use the notation $\ReadL (\D)$ for the multiset of names
that are objects of read actions. The functions $\WriteL {}$ and
$\ReadL {}$ may be inductively defined as follows: $\WriteL (\Write
a.\D) = \WriteL (\D) \uplus \{a\}$; $\ReadL (\Read a.\D) = \ReadL (\D)
\uplus \{a\}$; $\WriteL (\Read a.\D) = \ReadL (\Write a.\D) = \WriteL
(\D)$; and $\WriteL (\epsilon) = \ReadL (\epsilon) = \epsilon$.

\mysubsection{Example: Composing Synchronization.}
Before we describe the meaning of processes, we try to convey the
semantics of \AtCCS (and the usefulness of the atomic block operator)
using a simple example. We take the example of a concurrent system
with two memory cells, $M_1$ and $M_2$, used to store integers. We
consider here a straightforward extension of the calculus with
``value-passing\footnote{Keeping to our attention to economy in the
  definition of \AtCCS, we choose not to consider values in the formal
  syntax, but our results could be easily extended to take them into
  account.}.'' In this setting, we can model a cell with value $v$ by
an output $\Out {m_i}!v$ and model an update by a process of the form
$\Inp m_i?x .  (\Out {m_i}!v'
\parall \dots)$. With this encoding, the channel name $m_i$ acts as a
lock protecting the shared resource $M_i$.

Assume now that the values of the cells should be synchronized to
preserve a global invariant on the system. For instance, we model a
flying aircraft, each cell store the pitch of an aileron and we need
to ensure that the aileron stay aligned (that the values of the cells
are equal). A process testing the validity of the invariant is for
example $P_1$ below (we suppose that a message on the reserved channel
$\mathit{err}$ triggers an alarm). There are multiple design choices
for resetting the value of both cells to $0$, e.g. $P_2$ and $P_3$.
\[
\begin{array}{c}
  P_1 \ \defi\  \Inp {m_1}?x . \Inp {m_2}?y .  \text{if } x
  \mathop{\text{!=}} y \ \text{then}\, \Out {\mathit{err}}!\\
  P_2 \ \defi\ \Inp {m_2}?x . \Inp {m_1}?y . \bigl( \Out {m_1} ! 0 
  \parall \Out {m_2} ! 0) \bigr) \qquad
  P_3 \ \defi\ \Inp {m_1}?x . \bigl( \Out {m_1} ! 0 \parall
  \Inp {m_2}?y .  \Out {m_2} ! 0 \bigr)
\end{array}
\]

Each choice exemplify a problem with lock-based programming. The
composition of $P_1$ with $P_2$ leads to a race condition where $P_1$
acquire the lock on $M_1$, $P_2$ on $M_2$ and each process gets stuck.
The composition of $P_1$ and $P_3$ may break the invariant (the value
of $M_1$ is updated too quickly). A solution in the first case is to
strengthen the invariant and enforce an order for acquiring locks, but
this solution is not viable in general and opens the door to
\emph{priority inversion} problems. Another solution is to use an
additional (master) lock to protect both cells, but this approach
obfuscate the code and significantly decreases the concurrency of the
system.

Overall, this simple example shows that synchronization constraints do
not compose well when using locks. This situation is consistently
observed (and bears a resemblance to the inheritance anomaly problem
found in concurrent object-oriented languages). The approach advocated
in this paper is to use atomic transactions. In our example, the
problem is solved by simply wrapping the two operations in a
transaction, like in the following process, which ensures that all
cell updates are effected atomically.
\[
\Atomic\bigl(\Read ({m_2}?y) .  \Write ({m_2}!0) .  \Read ({m_1}?x)
.  \Write ({m_1}!0) \bigr) \]
More examples may be found on the paper on composable memory
transactions~\cite{STM}, which makes a compelling case that ``even
correctly-implemented concurrency abstractions cannot be composed
together to form larger abstractions.''

\mysubsection{Operational Semantics.}
Like for the syntax, the semantics of \AtCCS is stratified in two
levels: there is one reduction relation for processes and a second for
atomic expressions. With a slight abuse of notation, we use the same
symbol ($\to$) for both relations.

{\renewcommand{\GAP}{1em}
  \begin{table}[!t]
    \centering{\small
      \framebox{
        $\begin{array}{rl@{\quad}rl}
          \rulen{(out)}&\Out a \pv \Si \to \nil \pv \Si \uplus \{a\} &
          \rulen{(rep)}& *\Inp a.P \pv \Si \uplus \{a\} \to P\parall * \Inp a.P \pv \Si \\[\GAP]
          \rulen{(in)}&\Inp a.P \pv \Si\uplus \{a\} \to P \pv \Si&
          \rulen{(com)}&\deduce{P \pv \Si\to P' \pv \Si\uplus \{a\}\quad Q \pv \Si\uplus
            \{a\}\to Q' \pv \Si}
          {P\,|\,Q\to P'\parall Q'}\\[1.5em]
          \rulen{(parL)}&\deduce{P \pv \Si\to P' \pv \Si'}{P
            \parall Q \pv \Si \to P' \parall Q \pv \Si'}&
          \rulen{(hid)}&\deduce{P \pv \Si\uplus \{a^n\}\to
            P' \pv \Si'\uplus\{a^m\}\qquad a\notin  \Si,\Si'}
          {P\Res{n} a \pv \Si \to  P'\Res{m} a \pv \Si'}\\[1.5em]
          \rulen{(parR)}&\deduce{Q \pv \Si \to Q' \pv \Si'}{P
            \parall Q \pv \Si \to P \parall Q' \pv \Si'}&
          \rulen{(atSt)}&\Atomic (M) \pv \Si \to \Ablock{\Eblock{M}{\Si;\epsilon}}{M} \pv \Si\\[1.5em]
          \rulen{(atPass)}&\deduce{A\tom  A'}{\Ablock{A}{M} \pv \Si\to \Ablock{A'}{M} \pv \Si}&
          \rulen{(atRe)}&{\Ablock{\Eblock{\Retry}{\Si';\D}}{M} \pv \Si\to  \Atomic(M) \pv \Si}\\[1.5em]
          \multicolumn{4}{c}{\rulen{(atFail)}\quad\deduce{\ReadL (\D)\nsubseteq\Si}
            {\Ablock{\Eblock{\End}{\Si';\D}}{M} \pv \Si\to  \Atomic(M) \pv \Si}}\\[1.5em]
          \multicolumn{4}{c}{\rulen{(atOk)}\quad\deduce{ \ReadL (\D)\subseteq\Si 
              \quad \Si = \Si'' \uplus \ReadL (\D) \quad
              \WriteL (\D) = \{a_1,\dots,a_n\}}{\Ablock{\Eblock{\End}{\Si';\D}}{M} \pv \Si\to 
              \Out{a_1} \parall \dots \parall \Out{a_n} \pv \Si''}}\\[\GAP]
        \end{array}$ }\vspace*{\GAP}} 
    \caption{Operational Semantics: Processes.}\label{operationalsemanticsprocesses}
  \end{table}}

\myparagraph{Reduction for Processes.}
Table~\ref{operationalsemanticsprocesses} gives the semantics of
processes. A reduction is of the form $P \pv \Si \to P' \pv \Si'$
where $\Si$ is the state of $P$. The state $\Si$ records the names of
all output actions visible to $P$ when reduction happens. It grows
when an output is reduced, \rulen{(out)}, and shrinks in the case of
inputs, \rulen{(in)} and \rulen{(rep)}. A parallel composition evolves
if one of the component evolves or if both can synchronize, rules
\rulen{(parL)}, \rulen{(parR)} and \rulen{(com)}. In a hiding
$P\Res{n}{a}$, the annotation $n$ is an integer denoting the number of
outputs on $a$ which are visible to $P$. Intuitively, in a
``configuration'' $P\Res{n}{a} \pv \Si$, the outputs visible to $P$
are those in $\Si \uplus \{ a^n \}$. This extra annotation is
necessary because the scope of $a$ is restricted to $P$, hence it is
not possible to have outputs on $a$ in the global state. Rule
\rulen{(hid)} allows synchronization on the name $a$ to happen inside
a hiding. For instance, we have $(P
\parall \Out a)\Res{n}a \pv \Si \to P \Res{n+1} a \pv \Si$.

The remaining reduction rules govern the evolution of atomic
transactions. Like in the case of \rulen{(com)}, all those rules, but
\rulen{(atOk)}, leave the global state unchanged. Rule \rulen{(atSt)}
deals with the initiation of an atomic block $\Atomic(M)$: an ongoing
block $\Ablock{\Eblock M {\Si ;\epsilon}}{M}$ is created which holds
the current evaluation state $\Si$ and an empty log $\epsilon$. An
atomic block $\Ablock{A}{M}$ reduces when its expression $A$ reduces,
rule \rulen{(atPass)}. (The reduction relation for ongoing expression
is defined by the rules in Table~\ref{operationalsemanticsmemory}.)
Rules \rulen{(atRe)}, \rulen{(atFail)} and \rulen{(atOk)} deal with
the completion of a transaction. After a finite number of transitions,
the evaluation of an ongoing expression will necessarily result in a
fail state, $\Eblock{\Retry}{\Si;\D}$, or a success,
$\Eblock{\End}{\Si;\D}$. In the first case, rule \rulen{(atRe)}, the
transaction is aborted and started again from scratch. In the second
case, we need to check if the log is consistent with the current
evaluation state. A log is consistent if the read actions of $\D$ can
be performed on the current state. If the check fails, rule
\rulen{(atFail)}, the transaction aborts. Otherwise, rule
\rulen{(atOk)}, we commit the transaction: the names in $\ReadL (\D)$
are taken from the current state and a bunch of outputs on the names
in $\WriteL (\D)$ are generated.

{\renewcommand{\GAP}{1em}
  \begin{table}[!t]
    \centering{\small
      \framebox{
        $\begin{array}{rl@{\quad}rl}
          \rulen{(ARdOk)}&\deduce{\ReadL (\D) \uplus \{a\}  \subseteq \Si}
          {\Eblock{\Read a.M}{\Si;\D}\tom \Eblock M{\Si;\D. \Read a}}&
          \rulen{(ARdF)}&\deduce{\ReadL (\D) \uplus \{a\} \nsubseteq \Si}
          {\Eblock{\Read a.M}{\Si;\D}\tom \Eblock{\Retry}{\Si;\D} }\\[1.5em]
          \multicolumn{4}{c}{\rulen{(AWr)}
            \quad{\Eblock{\Write a.M}{\Si;\D}\tom \Eblock M{\Si;\D. \Write a}}}\\[\GAP]
          \multicolumn{4}{c}{\rulen{(AOI)}\quad{\Eblock{ M_{1}\Orelse
                M_{2}}{\Si;\D}}
            \tom {\Eblock{M_{1}}{\Si;\D} }\Oorelse {\Eblock{M_{2}}{\Si;\D}}}\\[\GAP]
          \rulen{(AOF)}&{\Eblock{\Retry}{\Si;\D}\Oorelse B\tom  B}&
          \rulen{(AOE)}&{\Eblock{\End}{\Si;\D}\Oorelse B\tom \Eblock{\End}{\Si;\D} }\\[\GAP]
          \rulen{(AOL)}&\deduce{A\tom A'}{A \Oorelse B\tom  A' \Oorelse B}&
          \rulen{(AOR)}&\deduce{B\tom B'}{A \Oorelse B\tom A \Oorelse B'}\\[\GAP]
        \end{array}$ }\vspace*{\GAP}}
    \caption{Operational Semantics : Ongoing Atomic Expression.}\label{operationalsemanticsmemory}
  \end{table}}

\myparagraph{Reduction for Ongoing Expressions.}
Table~\ref{operationalsemanticsmemory} gives the semantics of ongoing
atomic expressions. We recall that, in an expression $\Eblock{\Read
  a.M}{\Si;\D}$, the subscript $\Si$ is the \emph{initial state}, that
is a copy of the state at the time the block has been created and $\D$
is the log of actions performed since the initiation of the
transaction.

Rule \rulen{(ARdOk)} states that a read action $\Read a$ is recorded
in the log $\D$ if all the read actions in $\D . \Read a$ can be
performed in the initial state. If it is not the case, the ongoing
expression fails, rule \rulen{(ARdF)}. This test may be interpreted as
a kind of optimization: if a transaction cannot commit in the initial
state then, should it commit at the end of the atomic block, it would
mean that the global state has been concurrently modified during the
execution of the transaction. Note that we consider the initial state
$\Si$ and not $\Si \uplus \WriteL (\D)$, which means that, in an
atomic block, write actions are not directly visible (they cannot be
consumed by a read action). This is coherent with the fact that
outputs on $\WriteL (\D)$ only take place after commit of the block.
Rule \rulen{(AWr)} states that a write action always succeeds and is
recorded in the current log.

The remaining rules govern the semantics of the $\Retry$, $\End$ and
$\ofont{orElse}$ constructs. These constructs are borrowed from the STM
combinators used in the implementation of an STM system in Concurrent
Haskell~\cite{STM}. We define these operators with an equivalent
semantics, with the difference that, in our case, a state is not a
snapshot of the (shared) memory but a multiset of visible outputs. A
composition $M \Orelse N$ corresponds to the interleaving of the
behaviors of $M$ and $N$, which are independently evaluated with
respect to the same evaluation state (but have distinct logs). The
$\ofont{orElse}$ operator is preemptive: the ongoing block $M \Orelse N$ ends
if and only $M$ ends or $M$ aborts and $N$ ends.

\section{Encoding Concurrency Primitives}
\label{sec:encod-conc-prim}
Our first example is a simple solution to the celebrated \emph{leader
  election} problem that does not yield to deadlock and ensures that,
at each round, a leader is elected.

Consider a system composed by $n$ processes and a token, named $t$,
that is modeled by an output $\Out t$. A process becomes a leader by
getting (making an input on) $t$. As usual, all participants run the
same process (except for the value of their identity). We suppose that
there is only one copy of the token in the system and that leadership
of process $i$ is communicated to the other processes by outputting on
a reserved name $\mathit{win}_{i}$. A participant that is not a leader
output on $\mathit{loose}_i$. The protocol followed by the
participants is defined by the following process:
\[
\begin{array}{rcl}
  L_i &\defi& \bigl(\Atomic\bigl(
  \Read t.\Write {k}.\End \Orelse \Write
  {k'}.\End \bigr) \parall \Inp k.(\Out{\mathit{win}_{i}} \parall 
  \Out t) \parall \Inp {k'}.\Out{\mathit{loose}_{i}}\bigr) \Res{0} k
  \Res{0} {k'}
\end{array}
\]

In this encoding, the atomic block is used to protect the concurrent
accesses to $t$. If the process $L_i$ commits its transaction and
inputs (grabs) the token, it immediately release an output on its
private channel $k$. The transactions of the other participants may
either fail or commit while releasing an output on their private
channel $k'$. Then, the elected process $L_i$ may proceed with a
synchronization on $k$ that triggers the output $\Out
{\mathit{win}_i}$ and release the lock. The semantics of $\Atomic(\,)$
ensures that only one transaction can acquire the lock and commit the
atomic block, then no other process have acquired the token in the
same round and we are guaranteed that there could be at most one
leader.

This expressivity result is mixed blessing. Indeed, it means that any
implementation of the atomic operator should be able to solve the
leader election problem, which is known to be very expensive in the
case of loosely-coupled systems or in presence of failures (see
e.g.~\cite{palam} for a discussion on the expressivity of process
calculi and electoral systems). On the other hand, atomic transactions
are optimistic and are compatible with the use of probabilistic
approaches. Therefore it is still reasonable to expect a practical
implementation of \AtCCS.

In the following, we show how to encode two fundamental concurrency
patterns, namely (preemptive versions of) the choice and join-pattern
operators.

\mysubsection{Guarded choice.} 
We consider an operator for choice, $\act_1.P_1+\cdots+\act_n.P_n$,
such that every process is prefixed by an action $\act_i$ that is
either an output $\Out a_i$ or an input $\Inp a_i$. The semantics of
choice is characterized by the following three reduction rules (we
assume that $Q$ is also a choice):
\[
\begin{array}{rl@{\qquad}rl}
  \rulen{(c-inp)} & a.P+Q \pv \Si\uplus\{a\} \to  P \pv \Si&
  \rulen{(c-out)} & \Out a.P+Q \pv \Si \to  P \pv \Si \uplus \{a\} \\[\GAP]
  \multicolumn{4}{c}{\rulen{(c-pass)} \quad 
    \deduce{a\notin\Si \quad Q \pv \Si\to  Q' \pv \Si'}{a.P+Q \pv \Si\to  Q' \pv \Si'}}
\end{array}
\]

A minor difference with the behavior of the choice operator found in
CCS is that our semantics gives precedence to the leftmost process
(this is reminiscent of the preemptive behavior of $\ofont{orElse}$). Another
characteristic is related to the asynchronous nature of the calculus,
see rule \rulen{(c-out)}: since an output action can always interact
with the environment, a choice $\Out a . P + Q$ may react at once and
release the process $\Out a \parall P$.

Like in the example of the leader election problem, we can encode a
choice $\act_1.P_1+\cdots+\act_n.P_n$ using an atomic block that
will mediate the interaction with the actions $\act_1, \dots,
\act_n$. We start by defining a straightforward encoding of input/output
actions into atomic actions: $\tradm {\Out a} = \Write a$ and $\tradm
{\Inp a} = \Read a$. Then the encoding of choice is the process
\[
\begin{array}{rcl} 
  \tradm{\Inp {\act_1}.P_1 + \cdots + \Inp {\act_n}.P_n}
  &\defi& \bigl( \Atomic\bigl( \tradm{\act_1} . \tradm{\Out k_1}.\End
  \ \Orelse\ \cdots\ \Orelse\ \tradm{\act_n} . \tradm{\Out k_n} . \End
  \bigr)\\
  & & \quad
  \,\parall\,
  \Inp k_1.\tradm{P_1}\parall \cdots \parall \Inp k_n.\tradm{P_n} 
  \bigr) \Res{0}{k_1} \dots \Res{0}{k_n}
\end{array}\]

The principle of the encoding is essentially the same that in our
solution to the leader election problem. Actually, using the encoding
for choice, we can rewrite our solution in the following form: $L_i
\defi \Inp t .  (\Out{\mathit{win}_{i}} \parall \Out t) +
\Out{\mathit{loose}_{i}} .  \nil$\,. Using the rules in
Table~\ref{operationalsemanticsprocesses}, it is easy to see that our
encoding of choice is compatible with rule \rulen{(c-inp)}, meaning
that:
\[
\begin{array}{rll}
  \tradm{\Inp a . P + Q} \pv \Si \uplus \{ a \} & \to^* & \bigl(
  \Ablock{\Eblock{\End}{\Si \uplus \{ a \} ; \Read a . \Write {k_1}}}{M} \parall \Inp
  k_1.\tradm{P} \parall \dots \bigr) \Res{0}{k_1} \Res{}{\dots} \pv \Si
  \uplus \{ a \}\\
  & \to & \bigl(
  \Out {k_1} \parall \Inp
  k_1.\tradm{P} \parall \dots \bigr) \Res{0}{k_1} \Res{}{\dots} \pv \Si\\
  & \to & \bigl( \tradm{P} \parall \dots \bigr) \Res{0}{k_1} \Res{}{\dots} \pv \Si
\end{array}
\]
where the processes in parallel with $\tradm{P}$ are harmless.  In the
next section, we define a weak bisimulation equivalence $\Abisim$ that
can be used to garbage collect harmless processes in the sense that,
e.g. $(P \parall \Inp k . Q) \Res{0}{k} \Abisim P$ if $P$ has no
occurrences of $k$.  Hence, we could prove that $\tradm{\Inp a . P +
  Q} \pv \Si \uplus \{ a \} \to^* \Abisim \tradm{P} \pv \Si$, which is
enough to show that our encoding is correct with respect to rule
\rulen{(c-inp)}.  The same is true for rules \rulen{(c-out)} and
\rulen{(c-pass)}.

\mysubsection{Join Patterns.}
A multi-synchronization $(a_{1} \times\cdots\times a_{n}) .  P$ may be
viewed as an extension of input prefix in which communication requires
a synchronization with the $n$ outputs $\Out {a_1}, \dots, \Out {a_n}$
at once. that is, we have the reduction:
\[ 
\rulen{(j-inp)} \quad (a_{1} \times\cdots\times
a_{n}) . P \pv \Si \uplus \{a_1, \dots, a_n\} \ \to \
P \pv \Si \]

This synchronization primitive is fundamental to the definition of the
Gamma calculus of Ban\^atre and Le M\'etayer and of the Join calculus
of Fournet and Gonthier. It is easy to see that the encoding of a
multi-synchronization (input) is a simple transaction:
\[
\tradm{\bigl(a_1 \times \dots \times a_{n}\bigr) .  P} \ \defi\ \bigl(
\Atomic(\tradm {a_{1}}.\cdots.\tradm {a_{n}}.\tradm {\Out k}.\End)
\parall k .  \tradm{P} \bigr) \Res{0}{k} \quad \text{(where $k$ is
  fresh)}
\]
and that we have $\tradm{\bigl(a_1 \times \dots \times a_{n}\bigr) .
  P} \pv \Si \uplus \{a_1, \dots, a_n\} \ \to^* \bigl(\nil
\parall \tradm{P}\bigr) \Res {0}{k} \pv \Si$, where the process
$\bigl(\nil
\parall \tradm{P}\bigr) \Res {0}{k}$ is behaviorally equivalent to
$\tradm{P}$, that is:
\[
\tradm{\bigl(a_1 \times \dots \times a_{n}\bigr) .  P} \pv \Si \uplus
\{a_1, \dots, a_n\} \,\ \to^* \Abisim\ \, \tradm{P} \pv \Si
\]

Based on this encoding, we can define two interesting derived
operators: a mixed version of multi-synchronization, $(\act_{1}
\times\cdots\times \act_{n}) .  P$, that mixes input and output
actions; and a replicated version, that is analogous to replicated
input.
\[
\begin{array}{rcl}
  \tradm{\bigl(\act_1 \times \dots \times \act_{n}\bigr)
    .  P} &\ \defi\ & \bigl(
  \Atomic(\tradm{\act_{1}}.\cdots.\tradm{\act_{n}}.\tradm{\Out k}.\End)
  \parall k . \tradm{P} \bigr) \Res{0}{k}\\

  \tradm{*\, \bigl(\act_1 \times \dots \times \act_{n}\bigr)
    .  P} &\ \defi\ & \bigl( \Out r \ \parall
  * \Inp r . \Atomic(\tradm{\act_{1}}.\cdots.\tradm{\act_{n}}.\tradm{\Out
    r} . \tradm{\Out k}.\End)\ \parall * \Inp k . \tradm{P} \bigr) \Res{0}{r} \Res{0}{k}\\
\end{array}
\]

By looking at the possible reductions of these (derived) operators, we
can define derived reduction rules.  Assume $\D$ is the log
$\tradm{\act_1} .  \cdots . \tradm{\act_n}$, we have a simulation
result comparable to the case for multi-synchronization, namely:
\[
\begin{array}{rcl}
  \tradm{\bigl(\act_1 \times \dots \times \act_{n}\bigr)
    .  P} \pv \Si \uplus \ReadL(\D) &\ \to^* \Abisim\ & 
  \tradm{P} \pv \Si \uplus  \WriteL (\D)\\

  \tradm{* \bigl(\act_1 \times \dots \times \act_{n}\bigr)
    .  P}  \pv \Si \uplus \ReadL(\D) &\ \to^* \Abisim\ & 
  \tradm{* \bigl(\act_1 \times \dots \times \act_{n}\bigr)
    .  P} \ \parall\ \tradm{P} \pv \Si \uplus \WriteL (\D)\\
\end{array}
\]

To obtain join-definitions, we only need to combine a sequence of
replicated multi-synchronizations using the choice composition defined
precedently.  (We also need hiding to close the scope of the
definition.) Actually, we can encode even more flexible constructs
mixing choice and join-patterns. For the sake of simplicity, we only
study examples of such operations. The first example is the (linear)
join-pattern $(a \times b) . P \wedge (a \times c) . Q$, that may fire
$P$ if the outputs $\{a , b \}$ are in the global state $\Si$ and
otherwise fire $Q$ if $\{ a , c\}$ is in $\Si$ (actually, real
implementations of join-calculus have a preemptive semantics for
pattern synchronization). The second example is the derived operator
$(a\times b) + (b \times c \times \Out a).P$, such that $P$ is fired
if outputs on $\{ a , b\}$ are available or if outputs on $\{ b, c\}$
are available (in which case an output on $a$ is also generated).
These examples can be easily interpreted using atomic transactions:
\[
\begin{array}{rcll}
  \tradm{(a \times b) . P \wedge (a \times c) . Q} &\ \defi\ &
  \big( \Atomic\bigl( & \tradm{a}.\tradm{b}.\tradm{\Out k_1}.\End \Orelse\\
  &&& \tradm{a}.\tradm{c}.\tradm{\Out k_2}.\End \, \bigr)
  \ \parall\ \Inp k_1.P \ \parall\ \Inp k_2.Q \big) \Res{0} {k_1}\Res{0} {k_2}\\

  \tradm{\bigl( a\times b + b \times c \times \Out a \bigr).P} &\
  \defi\ &
  \big( \Atomic\bigl( & \tradm{a}.\tradm{b}.\tradm{\Out k}.\End \Orelse\\
  &&& \tradm{b}.\tradm{c}.\tradm{\Out a}.\tradm{\Out k}.\End \, \bigr)
  \ \parall\ \Inp k.P \big) \Res{0} {k}
\end{array}\]

In the next section we define the notion of bisimulation used for
reasoning on the soundness of our encodings. We also define an
equivalence relation for atomic expressions that is useful for
reasoning on the behavior of atomic blocks.

\section{Bisimulation Semantics}\label{sec:bisimulation}

A first phase before obtaining a bisimulation equivalence is to define
a Labeled Transition System (LTS) for \AtCCS processes related to the
reduction semantics.

\mysubsection{Labeled Semantics of \AtCCS.}
It is easy to derive labels from the reduction semantics given in
Table~\ref{operationalsemanticsprocesses}. For instance, a reduction
of the form $P \pv \Si \to P' \pv \Si \uplus \{ a \}$ is clearly an
\emph{output transition} and we could denote it using the transition
$P \tolabel{\Out a} P'$, meaning that the effect of the transition is
to add a message on $a$ to the global state $\Si$. We formalize the
notion of label and transition. Besides output actions $\Out a$, which
corresponds to an application of rule \rulen{(out)}, we also need
\emph{block actions}, which are multiset $\{ a_1, \dots, a_n \}$
corresponding to the commit of an atomic block, that is to the
deletion of a bunch of names from the global state in rule
\rulen{(atOk)}. Block actions include the usual labels found in LTS
for CCS and are used for labeling input and communication transitions:
an input actions $a$, which intuitively corresponds to rules
\rulen{(in)} and \rulen{(rep)}, is a shorthand for the (singleton)
block action $\{ a \}$; the silent action $\tau$, which corresponds to
rule \rulen{(com)}, is a shorthand for the empty block action
$\emptyset$. In the following, we use the symbols $\theta, \gamma,
\dots$ to range over block actions and $\mu, \mu', \dots$ to range
over labels, $\mu \, ::= \, \Out a \,\Alt\, \theta \,\Alt\, \tau
\,\Alt\, a$\,.

The labeled semantics for \AtCCS is the smallest relation $P
\tolabel{\mu} P'$ satisfying the two following clauses:
\begin{enumerate}
\item we have $P\tolabel{\Out a} P'$ if there is a state $\Si$ such
  that $P \pv \Si \to P' \pv \Si \uplus \{a\}$;
\item we have $P\tolabel{\theta} P'$ if there is a state $\Si$ such
  that $P \pv \Si \uplus \theta \to P' \pv \Si$.
\end{enumerate}

Note that, in the case of the (derived) action $\tau$, we obtain from
clause $2$ that $P\tolabel\tau P'$ if there is a state $\Si$ such that
$P \pv \Si\to P' \pv \Si$. As usual, silent actions label transitions
that do not modify the environment (in our case the global state) and
so are invisible to an outside observer. Unlike CCS, the calculus has
more examples of silent transition than mere internal synchronization,
e.g. the initiation and evolution of an atomic block, see e.g. rules
\rulen{(atST)} and \rulen{(atPass)}. Consequently, a suitable (weak)
equivalence for \AtCCS should not distinguish e.g. the processes
$\Atomic(\Retry)$, $\Atomic(\End)$, $(\Inp a . \Out a)$ and $\nil$.
The same is true with input transitions. For instance, we expect to
equate the processes $a.\nil$ and $\Atomic ({\Read a . \End})$.

Our labeled semantics for \AtCCS is not based on a set of transition
rules, as it is usually the case. Nonetheless, we can recover an
axiomatic presentation of the semantics using the tight correspondence
between labeled transitions and reductions characterized by
Proposition~\ref{prop:bisim-semant}.

\begin{proposition}\label{prop:bisim-semant}
  Consider two processes $P$ and $Q$. The following implications are
  true:
  \begin{description}
  \item[\rulen{(com)}] if $P \tolabel a P'$ and $Q \tolabel{\Out a}Q'$
    then $P \parall Q \tolabel\tau P' \parall Q'$;
  \item[\rulen{(par)}] if $P\tolabel \mu P'$ then $P\parall Q \tolabel
\mu P'\parall Q$ and $Q\parall P\tolabel \mu Q\parall P'$;
  \item[\rulen{(hid)}] if $P\tolabel \mu P'$ and the name $a$ does not
appear in $\mu$ then $P\Res n a \tolabel \mu P'\Res n a$;
  \item[\rulen{(hidOut)}] if $P\tolabel {\Out a} P'$ then $P\Res n a
\tolabel \tau P' \Res {n+1} a$;
\item[\rulen{(hidAt)}] if $P\tolabel \mu P'$ and $\mu= \theta \uplus
  \{a^{m}\}$, where $a$ is a name that does not appear in the label
  $\theta$, then $P\Res {n+m} a \tolabel\theta P' \Res {n} a$.
  \end{description}
\end{proposition}

\begin{proof} In each case, we have a transition of the form $P
  \tolabel \mu P'$.  By definition, there are states $\Si$ and $\Si'$
  such that $P \pv \Si \to P' \pv \Si'$. The property is obtained by a
  simple induction on this reduction (a case analysis on the last
  reduction rule is enough).\qed
\end{proof}

We define additional transition relations used in the remainder of the
paper. As usual, we denote by $\To$ the \emph{weak transition
  relation}, that is the reflexive and transitive closure of
$\tolabel{\tau}$. We denote by $\Tolabel{\mu}$ the relation $\To \,
\tolabel{\mu} \,\To$.  If $s$ is a sequence of labels $\mu_0 \dots
\mu_n$, we denote $\tolabel{s}$ the relation such that $P \tolabel{s}
P'$ if and only if there is a process $Q$ such that $P \tolabel{\mu_0}
Q$ and $Q \tolabel{\mu_1 \dots \mu_n} P'$ (and $\tolabel{s}$ is the
identity relation when $s$ is the empty sequence $\epsilon$). We also
define a weak version $\Tolabel s$ of this relation in the same way.
Lastly, we denote $\tolabel{a^{n}}$ the relation $\tolabel{\Inp a} \,
\dots \, \tolabel{\Inp a}$, the composition of $n$ copies of
$\tolabel{\Inp a}$.

\mysubsection{Asynchronous Bisimulation for Processes and Expressions.}
Equipped with a labeled transition system, we can define the
traditional (weak) bisimulation equivalence $\bisim$ between
processes. This is the largest equivalence $\mathcal R$ such that if
$P {\mathcal R} Q$ and $P \tolabel{\mu} P'$ then $Q \Tolabel{\mu} Q'$
and $P' {\mathcal R} Q'$. Weak bisimulation can be used to prove
interesting equivalences between processes. For instance, we can prove
that $(P \parall \Out a) \Res{n} a \bisim P \Res{n+1} a$. Nonetheless,
a series of equivalence laws are not valid for $\bisim$.  For
instance, $\Atomic(\Read a.\End ) \nbisim \Inp a .  \nil$, meaning
that, whereas there are no context that separates the two processes,
it is possible to test the presence of an atomic block.  Also, the
usual \emph{asynchronous law} is not valid: $\Inp a . \Out a \nbisim
\nil$.  To overcome these limitations, we define a weak
\emph{asynchronous bisimulation} relation, denoted $\Abisim$, in the
style of~\cite{ACS}.
\begin{definition}[weak asynchronous bisimulation]\label{def:bisim}
  A symmetric relation $\mathcal R$ is a weak asynchronous
  bisimulation if whenever $P\mathcal R Q$ then the following holds:
  \begin{enumerate}
  \item if $P\tolabel{\Out a} P'$ then there is $Q'$ such that
    $Q\Tolabel{\Out a}Q'$ and $P'\mathcal R Q'$;
  \item if $P\tolabel\theta P'$ then there is a process $Q'$ and a
    block action $\gamma$ such that $Q\Tolabel{\gamma} Q'$ and
    $\big(P' \parall \prod_{a\in{(\gamma \setminus \theta)}}\Out{a} \big)
    \mathop{\mathcal R} \big(Q'\parall \prod_{a\in{(\theta \setminus
        \gamma)}}\Out{a}\big)$.
  \end{enumerate}
  We denote with $\Abisim$ the largest weak asynchronous bisimulation.
\end{definition}

Assume $P \Abisim Q$ and $P \tolabel\tau P'$, the (derived) case for
silent action entails that there is $Q'$ and $\theta$ such that
$Q\Tolabel{\theta} Q'$ and $P' \parall \prod_{a\in \theta} \Out{a}
\Abisim Q'$. If $\theta$ is the silent action, $\theta = \{\,\}$, we
recover the usual condition for bisimulation, that is $Q \Tolabel{}
Q'$ and $P' \Abisim Q'$. If $\theta$ is an input action, $\theta = \{
a \}$, we recover the definition of asynchronous bisimulation
of~\cite{ACS}.  Due to the presence of block actions $\gamma$, the
definition of $\Abisim$ is slightly more complicated than
in~\cite{ACS}, but it is also more compact (we only have two cases)
and more symmetric. Hence, we expect to be able to reuse known methods
and tools for proving the equivalence of \AtCCS processes. Another
indication that $\Abisim$ is a good choice for reasoning about
processes is that it is a congruence. The proof is reported in Appendix~\ref{app:thcong}.
\begin{theorem}\label{bisimcong}
  Weak asynchronous bisimulation $\Abisim$ is a congruence.
\end{theorem}


We need to define a specific equivalence relation to reason on
transactions. Indeed, the obvious choice that equates two expressions
$M$ and $N$ if $\Atomic(M) \Abisim \Atomic(N)$ does not lead to a
congruence. For instance, we have $(\Read a .\Write a .\End)$
equivalent to $\End$ while $\Atomic (\Read a .\Write a .\End \Orelse
\Write b.\End) \ \nAbisim\ \Atomic (\End \Orelse \Write b.\End)$\,.
The first transaction may output a message on $b$ while the second
always end silently.

We define an equivalence relation between atomic expressions
$\AAbisim$, and a \emph{weak atomic preorder} $\AAmore$, that relates
two expressions if they end (or abort) for the same states. We also
ask that equivalent expressions should perform the same changes on the
global state when they end. We say that two logs $\D, \D'$ have same
effects, denoted $\D =_\Si \D'$ if $\Si \setminus \ReadL (\D) \uplus
\WriteL (\D) = \Si \setminus \ReadL (\D') \uplus \WriteL (\D')$. We
say that $M \AAmore_\Si N$ if and only if either (1)
$\Eblock{N}{\Si;\epsilon} \To \Eblock{\Retry}{\Si,\D}$; or (2)
$\Eblock{N}{\Si;\epsilon} \To \Eblock{\End}{\Si,\D}$ and
$\Eblock{M}{\Si;\epsilon} \To \Eblock{\End}{\Si;\D'}$. Similarly, we
have $M \AAbisim_\Si N$ if and only if either (1)
$\Eblock{M}{\Si;\epsilon} \To \Eblock{\Retry}{\Si,\D}$ and
$\Eblock{N}{\Si;\epsilon} \To \Eblock{\Retry}{\Si,\D'}$; or (2)
$\Eblock{M}{\Si;\epsilon} \To \Eblock{\End}{\Si;\D}$ and
$\Eblock{N}{\Si;\epsilon} \To \Eblock{\End}{\Si,\D'}$ with $\D
=_\sigma \D'$.

\begin{definition}[weak atomic equivalence]
  Two atomic expressions $M, N$ are equivalent, denoted $M\AAbisim N$,
  if and only if $M \AAbisim_\Si N$ for every state $\Si$. Similarly,
  we have $M \AAmore N$ if and only if $M \AAmore_\Si N$ for every
  state $\Si$.
\end{definition}

While the definition of $\AAmore$ and $\AAbisim$ depend on a universal
quantification over states, testing the equivalence of two expressions
is not expensive. First, we can rely on a monotonicity property of
reduction: if $\Si \subseteq \Si'$ then for all $M$ the effect of
$\Eblock{M}{\Si,\D}$ is included in those of $\Eblock{M}{\Si',\D}$.
Moreover, we define a normal form for expressions later in this
section (see Proposition~\ref{prop:normal-form}) that greatly
simplifies the comparison of expressions. Another indication that
$\AAbisim$ is a good choice of equivalence for atomic expressions is
that it is a congruence. The proof is reported in Appendix~\ref{app:thcong}.
\begin{theorem}\label{Abisimcong}
  Weak atomic equivalence $\AAbisim$ is a congruence.
\end{theorem}

\mysubsection{Dining Philosopher.}
In this example we give yet another solution to the well-known dining philosopher problem. We use atomic blocks of actions in the implementation of 
the system and we show that the obtained process behaves as its specification, without using backtracking and without falling into situations of deadlock. 
Suppose to have four philosophers,  $I=\{0,1,2,3\}$ is the considered set of indexes. In what follows we write $+$ for the sum modulo $4$. Suppose $t$ is a set of indexes corresponding to  thinking philosophers, 
which are ready to eat; and $e$  corresponds to eating philosophers, which are ready to think. $P_{t;e}$ is the specification of the system, with    
$t\cup e=I$, $t\cap e=\emptyset$ and there isn't $i\in I$ such that $i,i+1\in e$.
\[
\begin{array}{rcl}
P_{t;e} & \defi & \sum_{i\notin t} \Inp {t_{i}}.P_{t\cup i; e-i}\\
& + & \sum_{\{i=0,1 \textrm{ if }e=\emptyset\}} \tau. (\Inp {e_{i}}.P_{t-i; i}+ \Inp {e_{i+2}}.P_{t-(i+2);(i+2)})\\
& + & \sum_{\{i\in t\,|\, i-1,i+1\notin e,\, i+2\in e\}} \tau.\Inp{e_{i}}.P_{t-i;e\cup i}
\end{array}
\]
The system specification will never fall into deadlocks and there can be at most two eating philosophers (with indexes $i$ and $i+2$). The actions of eating and thinking 
of the philosopher $i$, $e_{i}$ and $t_{i}$, can be observed as inputs. 

A philosopher $D_{i}$, for $i\in I$, can be implemented as follows:
\[
D_{i}\defi\Atomic(\Read{c_{i-1}}.\Read{c_{i}}.\End).\Inp{e_{i}}.\Inp{t_{i}}.(\Out{c_{i-1}}\,|\,\Out{c_{i}}).
\]
Process $D_{i}$ attempts to get the chopsticks, on his right and left, by using  an atomic block for reading $c_{i-1}$ and $c_{i}$.
If the commit of the atomic block can not be performed then at least one of its neighbors, $D_{i-1}$ or $D_{i+1}$ is already eating, because at least one 
of the chopsticks is not available, thus $D_{i}$ will retry to get both chopsticks. 
Otherwise he can eat, thus he will acquire the chopsticks and eat by inputting $e_{i}$. After eating, he can decide to start thinking, 
thus he  reads $t_{i}$, and after that he releases both chopsticks.  

The global system is given by the parallel composition of the philosopher $D_{i}$ and the output of the four chopsticks, which are hidden to observers
\[
D\defi(D_{0}\,|\,D_{1}\,|\,D_{2}\,|\,D_{3}\,|\,\Out{c_{0}}\,|\,\Out{c_{1}}\,|\,\Out{c_{2}}\,|\,\Out{c_{3}})\Res {0}{c_{0},c_{1},c_{2},c_{3}}.
\]

In what follows we show that $ P_{I;\emptyset}\Abisim D$ holds. Before we need to define a useful abbreviation. 
Suppose $A,\,B,\,C,\,D,\,E\subseteq\{0,\,1,\,2,\,3\}$, are sets of indexes such that $A\cup B\cup C=\{0,\,1,\,2,\,3\}$, $A\cap B=A\cap C=B\cap C=\emptyset$ and 
$D\cup E\subseteq \{0,\,1,\,2,\,3\}$ with $D\cap E=\emptyset$. We define $\System A B C D E$ as follows:
\[\begin{array}{rcl}
\System A B C D E&\defi& (\prod_{\{i\in A\}}D_{i} \,|\,\prod_{\{i\in B\}} \Inp{e_{i}}.\Inp{t_{i}}.(\Out{c_{i-1}}\,|\,\Out{c_{i}})\\
&&\,|\,\prod_{\{i\in C\}} \Inp{t_{i}}.(\Out{c_{i-1}}\,|\,\Out{c_{i}})\\
&&\,|\,\prod_{\{i\in D\}}\Out{c_{i}})\Res {1}{c_{i}\,|\, i\in E} \Res {0}{c_{i}\,|\, i\in D}.
\end{array}
\]
That is a system where the philosophers in $A$ are in the initial state; philosophers in $B$ are ready to eat (they have already acquired the chopsticks); 
philosophers in $C$ are ready to think (they have already eaten); indexes in $D$ correspond to available chopsticks not yet outputted; indexes in $E$ correspond to 
chopsticks outputted, thus chopsticks that can be taken by some philosopher for eating.

In the following $\mathcal P(S)$ represents the powerset of $S$. $ P_{I;\emptyset}\mathcal R D$ where the weak asynchronous bisimulation $\mathcal R$  is defined as follows:
\[
\begin{array}{rcl}
\mathcal R & = & \big\{(P_{I;\emptyset},\System I \emptyset \emptyset {I\setminus S} S)\,|\,S\in \mathcal P(I)\big\}\\
& \cup & \big\{ (P_{I-i;i}, \System{I-i}{\emptyset}{\{i\}}{\{i+1,i+2\}\setminus S}{S})\,|\, S\in \mathcal P(\{i+1,i+2\})\big\}\\
& \cup & \big\{ (P_{I-i;i}, \System{\{i-1,i+1\}}{\{i+2\}}{\{i\}}{\emptyset}{\emptyset}) \big\}\\
& \cup & \big\{ (P_{\{i-1,i+1\};\{i,i+2\}}, \System{\{i-1,i+1\}}{\emptyset}{\{i,i+2\}}{\emptyset}{\emptyset})\big\}\\
& \cup & \big\{ ((\Inp{e_{i}}.P_{I-i;i}+\Inp{e_{i+2}}.P_{I-(i+2);(i+2)}), \System{\{i-1,i+1\}}{\{i+2,i\}}{\emptyset}{\emptyset}{\emptyset})\,|\,i=0,1\big\}\\
& \cup & \big\{ ((\Inp{e_{i}}.P_{I-i;i}+\Inp{e_{i+2}}.P_{I-(i+2);(i+2)}), \System{\{i-1,i,i+1\}}{\{i+2\}}{\emptyset}{\{i-1,i\}\setminus S}{S})\\
&& \qquad \qquad \,|\,S\in\mathcal P(\{i-1,i\}),\,i=0,1 \big\}\\
& \cup & \big\{ ((\Inp{e_{i}}.P_{I-i;i}+\Inp{e_{i+2}}.P_{I-(i+2);(i+2)}), \System{\{i-1,i+1,i+2\}}{\{i\}}{\emptyset}{\{i+1,i+2\}\setminus S}{S})\\
&&\qquad \qquad\,|\, S\in\mathcal P(\{i+1,i+2\}) ,\,i=0,1\big\}.
\end{array}\]

\begin{table}[!t]
  \centering{\small
    \framebox{
      $\begin{array}{lcrcl}
        \multicolumn{5}{l}{\text{Laws for atomic expressions:}}\\
        \quad & \rulen{(comm)} &\alpha.\beta.M&\ \AAbisim\ & \beta.\alpha.M\\
        & \rulen{(dist)}  &\alpha.(M\Orelse N)&\AAbisim & (\alpha.M)\Orelse (\alpha.N)\\
        & \rulen{(ass)}  &{M_1\Orelse (M_2\Orelse M_3)}&\AAbisim & {(M_1\Orelse M_2)\Orelse M_3}\\
        & \rulen{(idem)}  & M\Orelse M &\AAbisim & M\\
        & \rulen{(absRt1)}  &\alpha.\Retry & \AAbisim & \Retry\\
        & \rulen{(absRt2)}  &\Retry \Orelse M& \AAbisim & M \ \AAbisim\
        M \Orelse \Retry\\
        & \rulen{(absEnd)}  & \End \Orelse M & \AAbisim & \End\\
        \multicolumn{5}{l}{\text{Laws for processes:}}\\
        & \rulen{(asy)}  &\Inp a.\Out a &\ \Abisim\ &\nil \\
        & \rulen{(a-asy)}  &\Atomic(\Read a.\Write a.\End) & \Abisim &\nil \\
        & \rulen{(a-1)}  &\Atomic(\Read a.\End ) & \Abisim &\Inp a.\nil \\
      \end{array}$ }\vspace*{\GAP}} 
  \caption{Algebraic Laws of Transactions.}\label{table:lawM}
\end{table}

\mysubsection{On the Algebraic Structure of Transactions.}
The equivalence relations $\AAbisim$ and $\Abisim$ can be used to
prove interesting laws of atomic expressions and processes. We list
some of these laws in Table~\ref{table:lawM}. Appropriate bisimulation relations which prove laws in Table~\ref{table:lawM} are reported in Appendix~\ref{app:laws}. Let $\mathcal{M}$
denotes the set of all atomic expressions. The behavioral rules for
atomic expressions are particularly interesting since they exhibit a
rich algebraic structure for $\mathcal{M}$. For instance, rules
\rulen{(comm)} and \rulen{(dist)} state that action prefix $\alpha.M$
is a commutative operation that distribute over $\ofont{orElse}$. We
also have that $(\mathcal{M}, \ofont{orElse}, \Retry)$ is an
idempotent semigroup with left identity $\Retry$, rules \rulen{(ass)},
\rulen{(absRt2)} and \rulen{(idem)}, and that $\End$ annihilates
$\mathcal{M}$, rule \rulen{(absEnd)}. Most of these laws appear
in~\cite{STM} but are not formally proved.

Actually, we can show that the structure of $\mathcal{M}$ is close to
that of a bound join-semilattice. We assume unary function symbols
$a(\,)$ and $\Out a(\,)$ for every name $a$ (a term $\Out a(M)$ is
intended to represent a prefix $\Write a.M$) and use the symbols
$\sqcup, \one, \zero$ instead of $\ofont{orElse}, \End, \Retry$. With
this presentation, the behavioral laws for atomic expression are
almost those of a semilattice. By definition of $\AAmore$, we have
that $M \sqcup M' \AAbisim M$ if and only if $M \AAmore M'$ \label{AAmore} and for
all $M, N$ we have $\one \AAmore M \sqcup N \AAmore M \AAmore \zero$.
\[
\begin{array}{lll}
  \mu(\mu'(M)) \ \AAbisim\  \mu'(\mu(M)) \qquad &
  \mu(M \sqcup N) \ \AAbisim\  \mu(M) \sqcup
  \mu(N)
  \qquad &  \mu(\zero) \ \AAbisim\  \zero\\
  \zero \sqcup M \ \AAbisim\  M \ \AAbisim\
  M \sqcup \zero &  \one \sqcup M \ \AAbisim\  \one\\
\end{array}
\]

It is possible to prove other behavioral laws to support our
interpretation of $\ofont{orElse}$ has a join. However some important
properties are missing, most notably, while $\sqcup$ is associative,
it is not commutative. For instance, $a(\Out b(\one)) \sqcup \one
\nAAbisim \one$ while $\one \AAbisim \one \sqcup a(\Out b(\one))$,
rule \rulen{(absEnd)}. This observation could help improve the design
of the transaction language: it will be interesting to enrich the
language so that we obtain a real lattice.

\mysubsection{Normal Form for Transactions.}
Next, we show that it is possible to rearrange an atomic expression
(using behavioral laws) to put it into a simple \emph{normal form}.
This procedure can be understood as a kind of compilation that
transform an expression $M$ into a simpler form.

Informally, an atomic expression $M$ is said to be in \emph{normal
  form} if it does not contain nested $\ofont{orElse}$ (all
occurrences are at top level) and if there are no redundant branches.
A redundant branch is a sequence of actions that will never be
executed. For instance, the read actions in $\Read a.\End$ are
included in $\Read a.\Read b.\End$, then the second branch in the
composition $\bigl(\Read a.\End\bigr) \Orelse \bigl(\Read a.\Read
b.\End\bigr)$ is redundant: obviously, if $\Read a.\End$ fails then
$\Read a.\Read b.\End$ cannot succeed. We overload the functions
defined on logs and write $\ReadL (M)$ for the (multiset of) names
occurring in read actions in $M$. We define $\WriteL (M)$ similarly.
In what follows, we abbreviate $(M_{1} \Orelse \dots \Orelse M_{n})$
with the expression $\bigsqcup_{i \in 1..n}M_{i}$. We say that an
expression $M$ is in \emph{normal form} if it is of the form
$\bigsqcup_{i \in 1..n}K_{i}$ where for all indexes $i,j \in 1..n$ we
have: (1) $K_{i}$ is a sequence of action prefixes $\alpha_{j_{1}} .
\dots .  \alpha_{j_{n_i}} . \End$; and (2) $\ReadL ({K_{i}})
\nsubseteq \ReadL ({K_{j}})$ for all $i < j$.  Condition (1) requires
the absence of nested $\ofont{orElse}$ and condition (2) prohibits
redundant branches (it also means that all branches, but the last one,
has a read action). The following proposition is proved in Appendix~\ref{app:normal-form}.
\begin{proposition}\label{prop:normal-form}
  For every expression $M$ there is a normal form $M'$ such that $M
  \AAbisim M'$.
\end{proposition}


Our choice of using bisimulation for reasoning about atomic
transactions may appear arbitrary. We have already debated over the
need to consider asynchronous bisimulation $\Abisim$ instead of (the
simple) bisimulation $\bisim$. In the next section, we study a testing
equivalence for \AtCCS, more particularly an asynchronous may testing
semantics~\cite{DNH84}.

\section{May-testing semantics}\label{sec:trace}

Using a testing equivalence instead of bisimulation is sometimes more
convenient. Nonetheless, testing equivalences have the drawback that
their definition depends on a universal quantification over
arbitrarily many processes. We define a may-testing equivalence for
\AtCCS and give an alternative characterization using a trace-based
equivalence. We also expose some shortcomings of may testing related
to the (folklore) fact that it cannot distinguish the points of choice
in a process. Actually, we define for every atomic block $\Atomic(M)$
a corresponding process without transactions (but using choice) that
is indistinguishable from $\Atomic(M)$. The results enunciated in this section are proved in Appendix~\ref{app:may}.

We define the notion of observers and successful computations. An
\emph{observer} $O$ is a particular type of process which does not
contain atomic blocks and that can perform a distinct output $\Out w$
(the success action). We denote \Obs the set of all observers. A
\emph{computation} from a process $P$ and an observer $O$ is a
sequence of transitions of the form $P \parall O = P_{0}
\parall O_{0} \tolabel\tau \dots \tolabel\tau P_{k}\,|\,O_{k}
\tolabel\tau\dots$, which is either infinite or of finite size, say
$n$, such that $P_{n} \parall O_{n}$ cannot evolve. A computation from
$P \parall O$ is \emph{successful} if there is an index $n$ such that
$O_{n}$ has a success action, that is $O_{n} \tolabel{\Out w}$. In this
case, we say that $P\;\mathit{may}\;O$. Two processes are may testing
equivalent if they have the same successful observers.

\begin{definition}[may-testing preorder]
  Given two processes $P$ and $Q$, we write $P\maypr Q$ if for every
  observer $O$ in \Obs we have $P\;\mathit{may}\;O $ implies
  $Q\;\mathit{may}\;O $. We use $\mayeq$ to denote the equivalence
  obtained as the kernel of the preorder $\maypr$.
\end{definition}

Universal quantification on observers make it difficult to work with
the operational definition of the may preorder.
Following~\cite{BDNP}, we study a trace-based characterization for our
calculus. The following preorder over traces will be used for defining
the alternative characterization of the may-testing preorder.

In our setting, a \emph{trace} $s$ is a sequence of actions $\mu_1
\dots \mu_n$. (We only consider output and block actions and leave
aside $\tau$ and input actions, which are derivable). We define a
preorder $\tob$ on traces as the smallest relation that satisfies the
following laws.
\[
\begin{array}{crcl@{\qquad}crcl}
  \rulen{(TO1)} & s_1\, s_2 &\ \tob\ & s_1\, \{a\}\, s_2& 
  \rulen{(TO2)} & s_1\, s_2\, \{a\} s_3 &\ \tob\ & s_1\, \{a\}\, s_2\, s_3 \\
  \rulen{(TO3)} & s_1\, s_2 & \tob &  s_1\, \{a\} \, {\Out a}\ s_2 &
  \rulen{(TO4)} &  \{ a_{1},\dots,a_{n} \} & \multicolumn{2}{l}{\
    {}_0\!\succeq \, \tob\ \{a_{1}\} \dots \{a_{n}\}}
\end{array}
\]

Following the terminology of~\cite{BDNP}, \rulen{(TO1)}, \rulen{(TO2)}
and \rulen{(TO3)} are the laws for \emph{deletion},
\emph{postponement} and \emph{annihilation} of input action.  We add
rule \rulen{(TO4)} which allows to substitute block actions with the
corresponding sequences of inputs. The simulation relation $\tot$ is
the reflexive and transitive closure of $\tob$.  The preorder $\tot$
is preserved by prefixing.  We can now define a preorder over
processes.

\begin{definition}[alternative preorder]
  For processes $P$ and $Q$, we set $P\Ll Q$ if for all weak
  transition $P\Tolabel s P'$ there is a trace $s'$ and a process $Q'$
  such that $s'\tot s$ and $Q\Tolabel {s'} Q'$.
\end{definition}

We now prove coincidence of $\Ll$ and $\maypr$. Some definitions and
preliminary results are needed. For every label $\mu$ we define the
complement $\Out \mu$ such that: the complement of an output action
${\Out a}$ is a block action $\{ a \}$ and the complement of a block
action ${\{a_1, \dots, a_n\}}$ is a trace $\Out a_1 \dots \Out a_n$.
For every trace $s = \mu_1 \dots \mu_n$, the cotrace $\Out s = \Out
\mu_1 \dots \Out \mu_n$ is obtained by concatenating the complements of
the actions in $s$. The following lemma relates the preorder $\tot$
with the operational semantics of processes.

\begin{lemma}\label{lemma:toreduction}
  Assume that $s' \tot s$ and $P\Tolabel{\Out s} P'$, then there is a
  process $P''$ such that $P\Tolabel{\Out {s}'} P''$.
\end{lemma}

The next step is to define a special class of observers. For every
trace $s$, we inductively define an observer $\observer(s) \in \Obs$
as follows:
\[
\observer(\epsilon)\defi \Out w,\quad \observer(\Out a\, s)\defi \Inp
a.\observer(s),\quad \observer(\{a_{1},\dots,a_{n}\}\, s)\defi \bigl(
\prod_{i \in 1..n} \Out{a_i} \bigr) \parall \observer(s)
\]

The following property shows that the sequence of visible actions from
$\observer(s)$ is related to traces simulated by $s$.

\begin{lemma}\label{lemma:observer}
  Consider two traces $s$ and $r$. If there is a process $Q$ such that
  $\observer(s) \Tolabel{\Out r} \, \Tolabel{\Out w} Q$ then $r\tot
  s$.
\end{lemma}


We can now prove a full abstraction theorem between may testing
$\maypr$ and the alternative preorder $\Ll$.

\begin{theorem}\label{th:coincidence}
  For all processes $P$ and $Q$, we have $P\maypr Q$ if and only if
  $P\Ll Q$.
\end{theorem}

Next, we show that may-testing semantics is not precise enough to tell
apart atomic transactions from sequences of input actions. We consider
an atomic expression $M$ in normal form. Assume $M = \bigsqcup_{i \in
  1..n}M_{i}$, the following lemma state that the observing behavior
of $M$ is obtained by considering, for every branch $K_i$, a
transition labeled by the block action containing $\ReadL(K_i)$
followed by output transitions on the names in $\WriteL(K_i)$.

\begin{lemma}\label{lemma:nfend}
  Assume $M = \bigsqcup_{i \in 1..n}K_{i}$ is an expression in normal
  form. For every index $i$ in $\{1,\dots,n\}$ we have $\Atomic(M) \pv
  \Si_{i} \to^* \Ablock{\Eblock{\End}{\sigma_{i} ; \delta}}M \pv
  \sigma_i$ where $\sigma_{i} = \ReadL {(K_{i})} = \ReadL(\delta)$ and
  $\WriteL(\delta) = \WriteL(K_{i})$.
\end{lemma}

As a corollary of Lemma~\ref{lemma:nfend}, we obtain that the possible
behavior of $\Atomic(M)$ can be described as $\Atomic(M)
\Tolabel{\Si_{i}} \prod_{b \in \WriteL(K_i)} \Out b$ for every $i \in
1..n$, where $\Si_{i}$ is the multiset $\ReadL {(K_{i})}$.

We now prove that for every atomic transaction $\Atomic(M)$ there is a
CCS process $\tradm M$ that is may-testing equivalent to $M$. By CCS
process, we intend a term of \AtCCS without atomic transaction that
may include occurrences of the choice operator $P + Q$. By
Proposition~\ref{prop:normal-form}, we can assume that $M$ is in
normal form, that is $M = \bigsqcup_{i \in 1..n}K_{i}$. The
interpretation of a sequence of actions $K = \alpha_1 . \dots .
\alpha_n . \End$ is the process $\tradm K = \Inp a_1 . \cdots . \Inp
a_k . \bigl ( \Out {b_1}
\parall \dots \parall \Out {b_l} \bigr)$ where $\{ a_{1},\dots,a_{k}
\} = \ReadL (K)$ and $\{ b_{1},\dots,b_{l} \} = \WriteL (K)$. (In
particular we have $\tradm{\End} = \nil$.) The translated of $M$,
denoted $\tradm M$, is the process $\tradm {K_1} + \dots + \tradm
{K_n}$. The following theorem proves that may-testing semantics is not
able to distinguish the behavior of an atomic process from the
behavior of its translation, which means that may-testing is blind to
the presence of transactions.

\begin{proposition}\label{th:may}
  For every expression $M$ in normal form we have $\Atomic(M) \mayeq
  \tradm M$.
\end{proposition}


We observe that a process $\tradm M$ is a choice between processes of
the form $\Inp a . P$ or $\bigl(\prod_{i\in I} \Out {b_i}\bigr)$.
Therefore, using internal choice and a slightly more convoluted
encoding, it is possible to use only input guarded choice $\Inp a . P
+ \Inp b . Q$ in place of full choice in the definition of $\tradm M$.

\section{Future and Related Works}\label{sec:conclusions}

There is a long history of works that try to formalize the notions of
transactions and atomicity, and a variety of approaches to tackle this
problem. We review some of these works that are the most related to
ours.

We can list several works that combine ACID transactions with process
calculi. Gorrieri et al~\cite{GMM90} have modeled concurrent systems
with atomic behaviors using an extension of CCS.  They use a two-level
transition systems (a high and a low level) where high actions are
decomposed into atomic sequences of low actions.  To enforce
isolation, atomic sequences must go into a special invisible state
during all their execution. Contrary to our model, this work does not
follow an optimistic approach: sequences are executed sequentially,
without interleaving with other actions, as though in a critical
section. Another related calculus is RCCS, a reversible version of
CCS~\cite{DK04,DK05} based on an earlier notion of process calculus
with backtracking~\cite{BPW93}. In RCCS,each process has access to a
log of its synchronization's history and may always wind back to a
previous state. This calculus guarantees the ACD properties of
transactions (isolation is meaningless since RCCS do not use a shared
memory model). Finally, a framework for specifying the semantics of
transactions in an object calculus is given in~\cite{VJWH04}. The
framework is parametrized by the definition of a transactional
mechanism and allows the study of multiple models, such as e.g. the
usual lock-based approach. In this work, STM is close to a model
called \emph{versioning semantics}. Like in our approach, this model
is based on the use of logs and is characterized by an optimistic
approach where log consistency is checked at commit time.  Fewer works
consider behavioral equivalences for transactions.  A foundational
work is~\cite{BCGO03}, that gives a theory of transactions specifying
atomicity, isolation and durability in the form of an equivalence
relation on processes, but it provides no formal proof system.

Linked to the upsurge of works on Web Services (and on long running
Web transactions), a larger body of works is concerned with
formalizing \emph{compensating transactions}. In this context, each
transactive block of actions is associated with a compensation (code)
that has to be run if a failure is detected. The purpose of
compensation is to undo most of the visible actions that have been
performed and, in this case, atomicity, isolation and durability are
obviously violated. We give a brief survey of works that formalize
compensable processes using process calculi. These works are of two
types: {(1)} \emph{interaction based
  compensation}~\cite{BLZ03,BMM04,webpi}, which are extensions of
process calculi (like $\pi$ or join-calculus) for describing
transactional choreographies where composition take place dynamically
and where each service describes its possible interactions and
compensations; {(2)} \emph{compensable flow
  composition}~\cite{BMM05,ccsp,BHF04,BFN05}, where ad hoc process
algebras are designed from scratch to describe the possible flow of
control among services. These calculi are oriented towards the
orchestration of services and service failures. This second approach
is also followed in~\cite{BCGHM,BPG05} where two frameworks for
composing transactional services are presented.


The study of \AtCCS is motivated by our objective to better understand
the semantics of the STM model. Obtaining a suitable behavioral
equivalence for atomic expression is a progress for the verification
of concurrent applications that use STM.  However, we can imagine
using our calculus for other purposes.  An interesting problem is to
develop an approach merging atomic and compensating transactions. A
first step in this direction is to enrich our language and allow the
parallel composition of atomic expressions and the nesting of
transactions. We are currently working on this problem.  Another area
for research stems from our observation (see
Section~\ref{sec:bisimulation}) that the algebraic structure of atomic
expressions is lacking interesting property. Indeed, it will be
interesting to enrich the language of expressions in order to obtain a
real lattice. The addition of a symmetric choice operator for atomic
expressions may be a solution, but it could introduce unwanted
nondeterminism in the evaluation of transactions.

\newpage

\begin{appendix}
\section{Proofs of Section~\ref{sec:bisimulation}}\label{app:thcong}
Before proving the validity of Theorem~\ref{bisimcong} and Theorem~\ref{Abisimcong}, it is necessary to present some preliminary results.

The following proposition reminds an important property of asynchronous calculi:  no behavior causally depends on the execution of output actions. 
Relation $\sim$ stands for the usual strong bisimulation relation (see e.g.~\cite{M82}).
\begin{proposition_a}\label{prop:out}
$P\tolabel{\overline a} P'$ implies $P\sim P'\,|\, \Out a$.
\end{proposition_a}
\begin{proof}
By observing that outputs are non-blocking actions, a suitable strong bisimulation can be defined.\qed
\end{proof}

As direct consequences of the previous proposition, we get the
results enunciated in the following lemma: 
(1) output actions can always be delayed and (2)
a diamond property involving outputs.
\begin{lemma_a}\label{lemma-red}
  Let $\mu$ be a generic action ($\mu::=\overline b\,|\, \T\,|\,
  \tau$):
\begin{enumerate}
\item\label{one}\label{two}\label{three} $P\tolabel{\overline a}\tolabel{\mu}P'  $ implies $ P\tolabel{\mu}\tolabel{\overline a}P'$ ; 
similarly $P\tolabel{\overline a}\Tolabel{\mu}P' $ implies $ P\Tolabel{\mu}\tolabel{\overline a}P'$;

\item\label{four} $P\tolabel{\overline a}P' $ and $ P\tolabel{\mu}P''$ imply that there is a $P'''$ such that  $P'\tolabel{\mu}P'''$ and $P''\tolabel{\overline a}P'''$; 
similarly $P\tolabel{\overline a}P' $ and $ P\Tolabel{\mu}P''$ imply that there is a $P'''$ such that  $P'\Tolabel{\mu}P'''$ and $P''\tolabel{\overline a}P'''$.
\end{enumerate}
\end{lemma_a}
\begin{proof}
By Proposition~\ref{prop:out}.\qed
\end{proof}

The following propositions enunciate two relevant properties of the hiding operator.
\begin{proposition_a}\label{prop:out-res}
$(P\,|\,\Out a)\Res n b \Abisim (P\Res n b\,|\, \Out a) $ if $a\neq b$.
\end{proposition_a}
\begin{proof}
By Proposition~\ref{prop:bisim-semant} \rulen{(hid)}, and definition of $\tolabel \mu$.
\qed
\end{proof}

\begin{proposition_a}\label{prop:res-out}
$(P\,|\,\Out a)\Res n a \Abisim P\Res {n+1} a$.
\end{proposition_a}
\begin{proof}
It is enough to note that $(P\,|\,\Out a)\Res n a\tolabel\tau P \Res {n+1} a$, Proposition~\ref{prop:bisim-semant} \rulen{(hidAt)}.
\qed
\end{proof}

In the following propositions we prove that $\Abisim$ and $\AAbisim$ are closed under contexts; as a consequence we obtain that both are congruences.

\begin{proposition_a}\label{prop:inp}
$P\Abisim Q$ implies $\forall a:\;\Inp a.P\Abisim \Inp a.Q$.
\end{proposition_a}
\begin{proof}
It is enough to show that the relation $\mathcal R=\Abisim\cup \{(\Inp a. P,\Inp a. Q)\}$ is a weak asynchronous bisimulation.\qed
\end{proof}

\begin{proposition_a}\label{prop:repinp}
$P\Abisim Q$ implies $\forall a:\;*\Inp a.P\Abisim *\Inp a.Q$.
\end{proposition_a}
\begin{proof}
It is enough to show that the relation 
\[
\mathcal R\,=\, \{(( \prod_i P_{i}^{n_{i}}\,|\, *\Inp a.P),(\prod_i Q_i^{n_{i}}\,|\, *\Inp a.Q) )  \;\big|\; n_{i}\geq 0,\; (P_{i},Q_{i})\in \Abisim\}
\]
where $P^{n}$ is a shorthand for the parallel composition of $n$ copies of $P$ and $\prod_i P_{i}$ stands for $P_{1}|\cdots |P_{n}|\cdots$, is a weak asynchronous bisimulation up to $\sim$.

The proof proceeds as usual, by showing that every transition of the left term can be matched by a transition of the right one (and vice-versa), and the pair composed by the arrival processes is in $\mathcal R$. The proof is straightforward by a simple case analysis of transitions, as defined in Proposition~\ref{prop:bisim-semant}. The most involved case is when a communication occurs between two subprocesses, let's say $P_j$ and $P_k$. Suppose $P_j\tolabel a P_j'$ and $P_k\tolabel {\overline a}P_k'$. 
This means that, by Proposition~\ref{prop:bisim-semant} \rulen{(com)}:
\[
( \prod_i P_{i}^{n_{i}}\,|\, *\Inp a.P)\tolabel\tau (\prod_{i\neq j,k}P_i^{n_i}\,|\,P_j^{n_j-1}\,|\,P_k^{n_k-1}\,|\, P_j'\, |\, P_k'\,|\, *\Inp a.P)=R_1~.
\]
By $P_k\Abisim Q_k$ we know that $Q_k\Tolabel{\overline a}Q_k'$ with $P_k'\Abisim Q_k'$. We distinguish the following cases for $Q_j$:
\begin{description}
\item[$Q_j\Tolabel a Q_j'$:] in this case $Q_j'\Abisim P_j'$ and, by Proposition~\ref{prop:bisim-semant} \rulen{(com)}:  
\[
( \prod_i Q_{i}^{n_{i}}\,|\, *\Inp a.Q)\Tolabel\tau (\prod_{i\neq j,k}Q_i^{n_i}\,|\,Q_j^{n_j-1}\,|\,Q_k^{n_k-1}\,|\, Q_j'\, |\, Q_k'\,|\, *\Inp a.Q)=R_2~
\]
and $(R_1,R_2)\in \mathcal R$ by definition of $\mathcal R$.

\item[$Q_j\Tolabel\T Q_j'$:] this means that, by Proposition~\ref{prop:bisim-semant} \rulen{(par)}: 
\[
(\prod_i Q_i^{n_{i}}\,|\, *\Inp a.Q)\Tolabel\T (\prod_{i\neq j} Q_i^{n_{i}}\,|\,Q_j^{n_j-1}\,|\, Q_j'\,|\, *\Inp a.Q)=R_2
\]
and we have to show that $R_1\,|\, \prod_{b\in\T}\Out b\Abisim R_2$. We distinguish two cases:
\begin{description}
\item[$a\in\T$:] from $P_j\Abisim Q_j$ we obtain that $P_j'\,|\,\prod_{b\in\T\setminus a}\Out b \Abisim Q_j'$. Moreover, remembering that $P_k'\Abisim Q_k'$, we have (by definition of $\mathcal R$):
\[
(\prod_{i\neq j,k}P_i^{n_i}\,|\,P_j^{n_j-1}\,|\,P_k^{n_k-1}\,|\,P_j'\,|\, \prod_{b\in\T\setminus a}\Out b \,|\, P_k' \,|\,*\Inp a .P)\mathcal R (\prod_{i\neq j,k}Q_i^{n_i}\,|\,Q_j^{n_j-1}\,|\,Q_k^{n_k-1}\,|\, Q_j'\, |\, Q_k'\,|\, *\Inp a.Q)
\]
but $\Out a\Abisim\Out a$, thus we also have (again by definition of $\mathcal R$)
\[
(\prod_{i\neq j,k}P_i^{n_i}\,|\,P_j^{n_j-1}\,|\,P_k^{n_k-1}\,|\,P_j'\,|\, \prod_{b\in\T\setminus a}\Out b \,|\, P_k' \,|\,\Out a\,|\,*\Inp a .P)\mathcal R (\prod_{i\neq j,k}Q_i^{n_i}\,|\,Q_j^{n_j-1}\,|\,Q_k^{n_k-1}\,|\, Q_j'\, |\, Q_k'\,|\,\Out a\,|\, *\Inp a.Q)
\]
by Proposition~\ref{prop:out}, $\Out a\, |\, Q_k'\sim Q_k$, thus 
\[
(\prod_{i\neq j,k}P_i^{n_i}\,|\,P_j^{n_j-1}\,|\,P_k^{n_k-1}\,|\,P_j'\,|\, \prod_{b\in\T}\Out b \,|\, P_k' \,|\,*\Inp a .P)\mathcal R\sim (\prod_{i\neq j}Q_i^{n_i}\,|\,Q_j^{n_j-1}\,|\, Q_j'\,|\, *\Inp a.Q)~.
\]

\item[$a\notin\T$:] from $P_j\Abisim Q_j$ we obtain that $P_j'\,|\,\prod_{b\in\T}\Out b \Abisim Q_j'\,|\,\Out a$. Moreover, remembering that $P_k'\Abisim Q_k'$, we have (by definition of $\mathcal R$):
\[
(\prod_{i\neq j,k}P_i^{n_i}\,|\,P_j^{n_j-1}\,|\,P_k^{n_k-1}\,|\,P_j'\,|\, \prod_{b\in\T}\Out b \,|\, P_k' \,|\,*\Inp a .P)\mathcal R (\prod_{i\neq j,k}Q_i^{n_i}\,|\,Q_j^{n_j-1}\,|\,Q_k^{n_k-1}\,|\, Q_j'\,|\,\Out a\, |\, Q_k'\,|\, *\Inp a.Q)
\]
by Proposition~\ref{prop:out}, $\Out a\, |\, Q_k'\sim Q_k$, thus
\[
(\prod_{i\neq j,k}P_i^{n_i}\,|\,P_j^{n_j-1}\,|\,P_k^{n_k-1}\,|\,P_j'\,|\, \prod_{b\in\T}\Out b \,|\, P_k' \,|\,*\Inp a .P)\mathcal R\sim (\prod_{i\neq j}Q_i^{n_i}\,|\,Q_j^{n_j-1}\,|\, Q_j'\,|\, *\Inp a.Q)~.
\]\qed
\end{description}
\end{description}
\end{proof}

\begin{proposition_a}\label{prop:par}
$P\Abisim Q$ implies $\forall R:\; P\,|\,R\Abisim Q\,|\,R$.
\end{proposition_a}

\begin{proof}
The proof proceeds by showing that the relation
\[\mathcal R=\{ (P\,|\,R, Q\,|\, R)\,|\,(P,Q)\in \Abisim \}\]
is a weak asynchronous bisimulation up to $\sim$.

Suppose $P\,|\,R\tolabel \mu S$; by applying Proposition~\ref{prop:bisim-semant}, we can distinguish the following cases obtained by applying Proposition~\ref{prop:bisim-semant} \rulen{(par)} or  \rulen{(com)}:

\begin{description}
\item[$R\tolabel\mu R'$:]  $S=P\,|\,R'$; by Proposition~\ref{prop:bisim-semant} \rulen{(par)}, $Q\,|\,R\tolabel\mu Q\,|\, R'$ and  $(P\,|\,R') \mathcal R(Q\,|\, R')$ by definition of $\mathcal R$;

\item[$P\tolabel{\overline a}P'$:] $\mu=\overline a$ and $S=P'\,|\,R$. By $P\Abisim Q$ we have $Q\Tolabel{\overline a}Q'$ with $P'\Abisim Q'$. By Proposition~\ref{prop:bisim-semant} \rulen{(par)}, $Q\,|\,R\Tolabel{\overline a} Q'\,|\, R$ and $(P'\,|\,R)\mathcal R(Q'\,|\, R)$ by definition of $\mathcal R$;

\item[$P\tolabel{\T}P'$:]  $\mu= \T$ and $S=P'\,|\,R$. By $P\Abisim Q$ we have $Q\Tolabel{\T'} Q'$ and $\big(P'\,|\,\prod_{a\in\T'\setminus \T}\Out a\big)\Abisim\big(Q'\,|\,\prod_{a\in\T\setminus \T'}\Out a\big)$. 

By Proposition~\ref{prop:bisim-semant} \rulen{(par)}, $Q\,|\, R\Tolabel{\T'} Q'\,|\, R$ and  $\big(P'\,|\,\prod_{a\in\T'\setminus \T}\Out a\,|\, R\big)\mathcal R \big(Q'\,|\,\prod_{a\in\T \setminus \T'}\Out a\,|\, R\big)$ follows from $\big(P'\,|\,\prod_{a\in\T'\setminus \T}\Out a\big)\Abisim\big(Q'\,|\,\prod_{a\in\T \setminus \T'}\Out a\big)$ and definition of $\mathcal R$;

\item[$P\tolabel{\overline a} P'$ and $R\tolabel{a}R'$:] $\mu=\tau$ and $S=P'\,|\,R'$. $P\Abisim Q$ implies $Q\Tolabel{\overline a}Q'$ and $P'\Abisim Q'$. 
By Proposition~\ref{prop:bisim-semant} \rulen{(com)},  $Q\,|\,R\To Q'\,|\, R'$ and, by definition of $\mathcal R$, $(P'\,|\,R')\mathcal R (Q'\,|\, R')$;

\item[$P\tolabel{\{a\}} P'$ and $R\tolabel{\overline a}R'$:] $\mu=\tau$ and $S=P'\,|\,R'$. $P\Abisim Q$ implies that $Q\Tolabel\T Q'$. We consider the following cases by distinguishing the possible values of $\T$:
\begin{description}
\item[$\T =\{a\}$:] in this case $P'\Abisim Q'$. By Proposition~\ref{prop:bisim-semant} \rulen{(com)},   $Q\,|\,R\To Q'\,|\, R'$  and, by definition of $\mathcal R$,  $(P'\,|\,R')\mathcal R (Q'\,|\, R')$;


\item[otherwise:] $Q\,|\,R\Tolabel\T Q'\,|\, R$ by Proposition~\ref{prop:bisim-semant} \rulen{(par)}; we have to prove that $P'\,|\,R'\,|\,\prod_{b\in\T}\Out b\bisim Q'\,|\, R$. We distinguish the following cases:
\begin{description} 
\item[$a\in\T$:] from $P\Abisim Q$ we obtain $P'\,|\,\prod_{b\in\T\setminus a}\Out b\Abisim Q'$, by definition of $\mathcal R$:
\[
 P'\,|\, \prod_{b\in\T\setminus a}\Out b \,|\, R\bisim Q'\,|\, R
\]
and by Proposition~\ref{prop:out}, $R\sim R'\,|\,\Out a$, thus
\[
P'\,|\, R'\,|\,\prod_{b\in\T}\Out b \sim\Abisim Q'\,|\, R~;
\] 

\item[$a\notin\T$:] from $P\Abisim Q$ we obtain $P'\,|\,\prod_{b\in\T}\Out b\Abisim Q'\,|\,\Out a$, by definition of $\mathcal R$:
\[
 P'\,|\, \prod_{b\in\T}\Out b \,|\, R'\Abisim Q'\,|\,\Out a\,|\, R'
\]
and by Proposition~\ref{prop:out}, $R\sim R'\,|\,\Out a$, thus
\[
P'\,|\, R'\,|\,\prod_{b\in\T}\Out b \Abisim\sim Q'\,|\, R~.
\] \qed
\end{description}

\end{description}
\end{description}
\end{proof}

\begin{proposition_a}\label{prop:res}
$P\Abisim Q$ implies $\forall a,n\geq0:\; P\Res n a\Abisim Q\Res n a$.
\end{proposition_a}
\begin{proof}
The proof proceeds by showing that the relation:
\[\mathcal R=\{(P_{i}\Res{n+i} a,Q_{j}\Res {n+j} a)\,|\, n\geq 0,\; (P,Q)\in\Abisim, \; P\tolabel{\overline a^{i}}P_{i},\; Q\tolabel{\overline a^j}Q_{j}\}\]
is a weak asynchronous bisimulation up to $\sim$.
We distinguish the following cases:
\begin{description}
\item[\rulen{(hid)}:]   $P_{i}\Res {n+i} a\tolabel{\mu} P_{i}'\Res {n+i} a$ is derived by $P_{i}\tolabel{\mu}P_{i}'$, if $a$ not appears in $\mu$. 
By Lemma~\ref{lemma-red}~(\ref{one}), $P\tolabel{\overline a^{i}}P_{i}\tolabel{\mu}P_{i}'$ implies 
$P\tolabel{\mu}P'\tolabel{\overline a^{ i}}P_{i}'$. From $P\Abisim Q$ we obtain $Q\Tolabel{\mu}Q'$ with $P'\Abisim Q'$ and by 
$Q\tolabel{\overline a^{ j}} Q_{j} $ and Lemma~\ref{lemma-red}~(\ref{four}), $Q'\tolabel{\overline{a}^{ j}}Q_{j}'$ and  
$Q_{j}\Tolabel{\mu}Q_{j}'$; by Proposition~\ref{prop:bisim-semant} \rulen{(hid)}, $Q_{j}\Res{n+j}a\Tolabel{\mu}Q_{j}'\Res{n+j} a$. 
Finally, $(P_{i}'\Res{n+i}a) \mathcal R( Q_{j}'\Res{n+j}a)$ because of $P'\Abisim Q'$, 
$P'\tolabel{\overline a^{ i}}P_{i}'$, $Q'\tolabel{\overline a^{ j}}Q_{j}'$ and definition of $\mathcal R$;

\item[\rulen{(hidAt)}:]   $P_{i}\Res {n+i} a\tolabel{\T} P_{i}'\Res {n'} a$ is derived by $P_{i}\tolabel{\T'}P_{i}'$ with $\T'=\T\uplus a^{m}$  and 
$n'=n+i-m$. By Lemma~\ref{lemma-red}~(\ref{two}), $P\tolabel{\overline a^{ i}}P_{i}\tolabel\T P_{i}'$ implies 
$P\tolabel{\T}P'\tolabel{\overline a^{ i}}P_{i}'$. By $P\Abisim Q$, $Q\Tolabel {\gamma'} Q'$ with 
$(P'\,|\, \prod_{b\in \gamma'\setminus\T'}\Out b) \Abisim (Q'\,|\, \prod_{b\in\T'\setminus\gamma'}\Out b)$. 
 Suppose $\gamma'=\gamma\uplus a^{m'}$ and, without loss of generality, that  $m'>m$. 
We can rewrite $P'\,|\, \prod_{b\in \gamma'\setminus\T'}\Out b$ as $ P'\,|\, \Out a ^{m'-m}\,|\, \prod_{b\in \gamma \setminus\T}\Out b$ and 
$Q'\,|\, \prod_{b\in\T'\setminus\gamma'}\Out b$ as $Q'\,|\, \prod_{b\in\T \setminus\gamma}\Out b$, thus 
\[
(P'\,|\, \Out a ^{m'-m}\,|\, \prod_{b\in \gamma \setminus\T}\Out b) \Abisim (Q'\,|\, \prod_{b\in\T \setminus\gamma}\Out b).
\] 
Moreover, by Lemma~\ref{lemma-red}~(\ref{four}), $Q\Tolabel {\gamma'} Q'$ and $Q\tolabel{\overline a^{ j}}Q_{j}$ imply
 $Q_{j}\Tolabel{\gamma'}Q_{j}'$ and $Q'\tolabel{\overline a^j}Q_{j}'$; by Proposition~\ref{prop:bisim-semant} \rulen{(hidAt)}, 
 $Q_{j}\Res {n+j}a \Tolabel\gamma Q_{j}'\Res{n+j-m'}a$. 

We have to relate the processes $P_{i}' \Res{n+i-m} a\,|\, \prod_{b\in \gamma \setminus\T}\Out b$ and $Q_{j}' \Res{n+j-m'} a\,|\, \prod_{b\in \T \setminus\gamma}\Out b$. 

By Proposition~\ref{prop:bisim-semant} \rulen{(hidOut)}, 
$(P'\,|\, \Out a ^{m'-m}\,|\, \prod_{b\in \gamma \setminus\T}\Out b)\Res{n-m'}a \tolabel{\tau
}(P_{i}'\,|\, \prod_{b\in \gamma \setminus\T}\Out b)\Res{n+i-m}a$ and 
$(Q'\,|\, \prod_{b\in\T \setminus\gamma}\Out b)\Res{n-m'}a\tolabel{\tau
}(Q_{j}'\,|\, \prod_{b\in\T \setminus\gamma}\Out b)\Res{n+j-m'}a$; thus from 
$(P'\,|\, \Out a ^{m'-m}\,|\, \prod_{b\in \gamma \setminus\T}\Out b) \Abisim (Q'\,|\, \prod_{b\in\T \setminus\gamma}\Out b)$ we obtain 
$\big((P_{i}'\,|\, \prod_{b\in \gamma \setminus\T}\Out b)\Res{n+i-m}a\big) \mathcal R \big((Q_{j}'\,|\, \prod_{b\in\T \setminus\gamma}\Out b)\Res{n+j-m'}a\big)$, that is  
$P_{i}' \Res{n+i-m} a\,|\, \prod_{b\in \gamma \setminus\T}\Out b\sim \mathcal R\sim Q_{j}' \Res{n+j-m'} a\,|\, \prod_{b\in \T \setminus\gamma}\Out b$, by Proposition~\ref{prop:out-res}.

\item[\rulen{(hidOut)}:]  $P_{i}\Res {n+i} a\tolabel{\tau} P_{i}'\Res {n+i+1} a$ is derived by $P_i\tolabel{\overline a}P_i'$; $P_i'=P_{i+1}$  and by definition of 
$\mathcal R$ we have  $(P_{i+1}\Res{n+i+1}a) \mathcal R (Q_{j}\Res{n+j}a)$.\qed
\end{description}
\end{proof}

\begin{proposition_a}\label{prop:Mcongpref}
Suppose $\alpha= \Read a $ or $\alpha= \Write a$. If $M\AAbisim N$ then $\alpha.M \AAbisim \alpha.N$.
\end{proposition_a}
\begin{proof}
Consider the case $\alpha=\Read a$. It suffices to show that $\mathcal R \subseteq \AAbisim$, where 
\[
\begin{array}{rcl}
\mathcal R&=&\{ (\Eblock{\Read a.M}{\Si;\epsilon},\Eblock{\Read a.N}{\Si;\epsilon}),\,  (\Eblock{\Retry}{\Si;\epsilon},\Eblock{\Retry}{\Si;\epsilon})\} \cup \\
&&\{(\Eblock{M'}{\Si;\Read a.\D}, \Eblock{N'}{\Si;\Read a.\D'})\,|\, (\Eblock{M'}{\Si \setminus\{a\};\D},\Eblock{N'}{\Si \setminus\{a\};\D'})\in \AAbisim, \\
&&\qquad\Eblock{M}{\Si \setminus\{a\};\epsilon}\To \Eblock{M'}{\Si \setminus\{a\};\D}\textrm{ and } \Eblock{N}{\Si \setminus\{a\};\epsilon}\To \Eblock{N'}{\Si \setminus\{a\};\D'}\}.
\end{array}
\]
Note that $M \AAbisim N$ implies $\D=_{\Si\setminus \{a\}}\D'$, thus $\Read a.\D =_{\Si}\Read a.\D'$.\qed
\end{proof}

\begin{proposition_a}\label{prop:Mcongorelse}
If $M_{1}\AAbisim N_{1}$ and $M_{2}\AAbisim N_{2}$ then $M_{1} \Orelse M_{2}\AAbisim N_{1}\Orelse N_{2}$.
\end{proposition_a}
\begin{proof}
It suffices to show that $\mathcal R \subseteq \AAbisim$, where 
\[
\begin{array}{rcl}
\mathcal R&=&\{ (\Eblock{M_{1} \Orelse M_{2}}{\Si;\epsilon},\Eblock{N_{1} \Orelse N_{2}}{\Si;\epsilon})\}\\ 
&\cup&\{((A\Orelse B),(C\Orelse D))\big | \Eblock{M_{1}}{\Si;\epsilon}\To A,\, \Eblock{M_{2}}{\Si;\epsilon}\To B,\, \Eblock{N_{1}}{\Si;\epsilon}\To C,\\
&&\qquad\qquad\qquad \Eblock{N_{2}}{\Si;\epsilon}\To D, (A,C)\in \AAbisim,\, (B,D)\in\AAbisim\} \\
&\cup &\{(B,D)\big|  \Eblock{M_{1}}{\Si';\epsilon}\To \Eblock{\Retry}{\Si';\delta},\, \Eblock{N_{1}}{\Si';\epsilon}\To \Eblock{\Retry}{\Si';\delta'},\, \Eblock{M_{2}}{\Si';\epsilon}\To B,\\
&&\qquad\qquad\qquad \Eblock{N_{2}}{\Si';\epsilon}\To D, (B,D)\in \AAbisim\}\\
&\cup&\{(\Eblock{\End}{\Si'';\D},\Eblock{\End}{\Si'';\D'})\,\big|\,\Eblock{M_{1}}{\Si'';\epsilon}\To\Eblock{\End}{\Si'';\D} ,\, \Eblock{N_{1}}{\Si'';\epsilon}\To\Eblock{\End}{\Si'';\D'}\}.
\end{array}
\]
Note that $M_{i} \AAbisim N_{i}$, for $i=1,2$, ensures that, in case of successful termination, the resulting logs have the same effects.\qed
\end{proof}


Weak atomic bisimulation entails weak asynchronous bisimulation, but the inverse does not hold. E.g. $\Atomic(\Read a .\Write a .\End)\Abisim \Atomic(\End)$ but $\Read a.\Write a.\End\not\AAbisim \End$.

\begin{proposition_a}\label{prop:abisim}
$M\AAbisim N$ implies $\Atomic(M)\Abisim \Atomic(N)$.
\end{proposition_a}
\begin{proof}
By contradiction, suppose that $\Atomic(M)\not\Abisim \Atomic(N)$. This means that there is a $\D$ such that $\Atomic(M)\Tolabel{\ReadL(\D)}P$, with 
$P=\prod _{b\in \WriteL(\D)}\Out b$,
and for every $\D'$ such that   $\Atomic(N)\Tolabel{\ReadL(\D')}Q$, with $Q=\prod _{b\in \WriteL(\D')}\Out b$, we have 
$(P\,|\, \prod _{b\in (\ReadL(\D')\setminus\ReadL(\D))}\Out b) \not\Abisim (Q\,|\, \prod _{b\in (\ReadL(\D)\setminus\ReadL(\D'))}\Out b)$.
This means that there is an $a$ such that $(P\,|\, \prod _{b\in (\ReadL(\D')\setminus\ReadL(\D))}\Out b) \tolabel{\overline a}$ and 
$ (Q\,|\, \prod _{b\in (\ReadL(\D)\setminus\ReadL(\D'))}\Out b)\not\tolabel{\overline a}$ (or vice versa).

By rules \rulen{(atPass)} and \rulen{(atOk)} and definition of $\tolabel\mu$,  $\Atomic(M)\Tolabel{\ReadL(\D)}P$ implies that there is a $\Si$ such that 
$\Eblock{M}{\Si;\epsilon}\To\Eblock{\End}{\Si;\D}\tolabel{\ReadL(\D)}P$. By definition of $\AAbisim$ there is a $\D''$ such that 
$\Eblock{N}{\Si;\epsilon}\To\Eblock{\End}{\Si;\D''}$, with $\D=_{\Si}\D''$, that is $\Si\setminus\ReadL(\D)\uplus\WriteL(\D)= \Si\setminus\ReadL(\D'')\uplus\WriteL(\D'')$. 
Thus by rules \rulen{(atPass)} and \rulen{(atOk)} and Proposition~\ref{prop:bisim-semant}  $\Atomic(N)\Tolabel{\ReadL(\D'')}Q$ with 
$Q=\prod _{b\in \WriteL(\D'')}\Out b$.

Suppose $P=\prod _{b\in \WriteL(\D)}\Out b\tolabel{\overline a}$; this means that $a\in \WriteL(\D)$. From  
$\Si\setminus\ReadL(\D)\uplus\WriteL(\D)= \Si\setminus\ReadL(\D'')\uplus\WriteL(\D'')$ we obtain $\WriteL(\D)= \WriteL(\D'')\uplus\ReadL(\D)\setminus\ReadL(\D'')$, hence or $Q=\prod _{b\in \WriteL(\D'')}\Out b\tolabel{\overline a}$ or 
$\prod _{b\in (\ReadL(\D)\setminus\ReadL(\D''))}\Out b\tolabel{\overline{a}}$. 

Suppose $a\in (\ReadL(\D'')\setminus\ReadL(\D))$, then $\WriteL(\D'')= \WriteL(\D)\uplus\ReadL(\D'')\setminus\ReadL(\D)$ implies that $a\in \WriteL(\D'')$, that is $Q\tolabel{\overline a}$.

In both cases we have a contradiction because we have assumed that $ (Q\,|\, \prod _{b\in (\ReadL(\D)\setminus\ReadL(\D''))}\Out b)\not\tolabel{\overline a}$. \qed
\end{proof}

We can now prove the main results of Section~\ref{sec:bisimulation}.

\begin{theorem_a}[Theorem~\ref{bisimcong}]
Weak asynchronous bisimulation $\Abisim$ is a congruence.
\end{theorem_a}
\begin{proof}
The result follows by Propositions~\ref{prop:inp}-\ref{prop:abisim}.\qed
\end{proof}

\begin{theorem_a}[Theorem~\ref{Abisimcong}]
Weak atomic bisimulation $\AAbisim$ is a congruence.
\end{theorem_a}
\begin{proof}
The result follows by Propositions~\ref{prop:Mcongpref} and~\ref{prop:Mcongorelse}.\qed
\end{proof}


\section{Proofs of laws in Table~\ref{table:lawM}}\label{app:laws}

Laws in Table~\ref{table:lawM} are proved, as usual, by showing  appropriate bisimulation relations. In the following cases $\mathcal R$ is the proposed bisimulation. In what follows $a\notin\Si$ means that the name $a$ does not appear in $\Si$ and $a^n\in\Si$ means that $\Si$ contains $n$ copies of $a$.

\begin{description}
\item[\rulen{(comm)} $\alpha.\alpha'.M\AAbisim  \alpha'.\alpha.M$:]  Suppose $\alpha=\Read a$ and $\alpha'=\Read b$ (the other cases are similar.) 
\[
\begin{array}{rcl}
\mathcal R & = &\big\{  ({\Eblock{\Read a.\Read b.M}{\Si;\epsilon}},{\Eblock{\Read b.\Read a.M}{\Si;\epsilon}})\big\}\\
&\cup&\{({\Eblock{\Read b.M}{\Si;\Read a}},  {\Eblock{\Read a.M}{\Si;\Read b}}), \,( {\Eblock{M''}{\Si;\Read a.\Read b.\D}},  {\Eblock{M''}{\Si;\Read b.\Read a.\D}})\\
&&\quad\big|  a^{n},b^{m} \in \Si,\:n,m>0 ,\:\Eblock{M}{\Si\setminus\{a,b\};\epsilon}\To \Eblock{M''}{\Si\setminus\{a,b\};\D}\}\\
&\cup&\big\{( {\Eblock{\Retry}{\Si;\epsilon}},  {\Eblock{\Retry}{\Si;\epsilon}})\big| a,b\notin \Si\}\\
&\cup&\big\{( {\Eblock{\Retry}{\Si;\epsilon}},  {\Eblock{\Read a.M}{\Si;\Read b}}), \,
( {\Eblock{\Retry}{\Si;\epsilon}},  {\Eblock{\Retry}{\Si;\Read b}}), \\
&&\quad \big| a\notin\Si,\:b^{m}\in\Si,\: m>0\big\}\\
&\cup&\big\{( {\Eblock{\Read b.M}{\Si;\Read a}},  {\Eblock{\Retry}{\Si;\epsilon}}), \,
( {\Eblock{\Retry}{\Si;\Read a}},  {\Eblock{\Retry}{\Si;\epsilon}}) \\
&&\quad \big| a^{n}\in\Si,\:b\notin\Si,\: n>0\big\}~.\\
\end{array}\]

\item[\rulen{(dist)} $\alpha.(M\Orelse N)\AAbisim  (\alpha.M)\Orelse (\alpha.N)$:]  Suppose $M'=\Read a.(M\Orelse N)$ and $N'=(\Read a.M) \Orelse (\Read a.N)$. 
\[
\begin{array}{rcl}
\mathcal R & = &\big\{( {\Eblock{M'}{\Si;\epsilon}}, {\Eblock{N'}{\Si;\epsilon}}),\,
( {\Eblock{M'}{\Si;\epsilon}}, {\Eblock{\Read a.M}{\Si;\epsilon}\Orelse \Eblock{\Read a.N}{\Si;\epsilon}})\big\}\\
&\cup&\big\{( {\Eblock{\Retry}{\Si;\epsilon}}, {\Eblock{\Retry}{\Si;\epsilon}\Orelse \Eblock{\Read a.N}{\Si;\epsilon}}),\\
&&( {\Eblock{\Retry}{\Si;\epsilon}}, {\Eblock{\Read a.M}{\Si;\epsilon}\Orelse \Eblock{\Retry}{\Si;\epsilon}}),\\
&&( {\Eblock{\Retry}{\Si;\epsilon}}, {\Eblock{\Retry}{\Si;\epsilon}\Orelse \Eblock{\Retry}{\Si;\epsilon}}),\,
( {\Eblock{\Retry}{\Si;\epsilon}}, {\Eblock{\Retry}{\Si;\epsilon}})\\
&&\quad \big| a\notin \Si \big\}\\
&\cup&\big\{( {\Eblock{M\Orelse N}{\Si;\Read a}}, {\Eblock{M}{\Si;\Read a}\Orelse \Eblock{N}{\Si;\Read a}}),\\
&&( {\Eblock{M'}{\Si;\epsilon}}, {\Eblock{M}{\Si;\Read a}\Orelse \Eblock{\Read a.N}{\Si;\epsilon}}),\,
( {\Eblock{M'}{\Si;\epsilon}}, {\Eblock{\Read a.M}{\Si;\epsilon}\Orelse \Eblock{N}{\Si;\Read a}})\\
&&\quad\big| a^n\in \Si,\: n>0\big\}\\
&\cup&\big\{( {A\Orelse \Eblock{N}{\Si;\Read a}}, {A\Orelse \Eblock{\Read a.N}{\Si;\epsilon}})\,|\, \Eblock{M}{\Si;\Read a}\To A,\: a^n\in \Si,\: n>0\big\}\\
&\cup&\big\{( {\Eblock{M}{\Si;\Read a}\Orelse B}, {\Eblock{\Read a.M}{\Si;\epsilon}\Orelse B})\,|\, \Eblock{N}{\Si;\Read a}\To B,\: a^n\in \Si,\: n>0\big\}\\
&\cup&\big\{( {A\Orelse B}, {A\Orelse B})\,|\, \Eblock{M}{\Si;\Read a}\To A,\, \Eblock{N}{\Si;\Read a}\To B,\: a^n\in \Si,\: n>0\big\}\\
&\cup&\big\{( {C}, {C})\,|\, \Eblock{M}{\Si;\Read a}\To \Eblock{\Retry}{\Si;\D},\, \Eblock{N}{\Si;\Read a}\To C,\: a^n\in \Si,\: n>0\big\}~.\\
\end{array}\]

\item[\rulen{(ass)} ${M_1\Orelse (M_2\Orelse M_3)}\AAbisim {(M_1\Orelse M_2)\Orelse M_3}$:]  
\[
\begin{array}{rcl}
\mathcal R & = &\big\{( {\Eblock{M_1\Orelse (M_2\Orelse M_3)}{\Si;\epsilon}}, {\Eblock{(M_1\Orelse M_2)\Orelse M_3}{\Si;\epsilon}}),\\
&& ( {\Eblock{M_{1}}{\Si;\epsilon}\Orelse \Eblock{M_{2}\Orelse M_{3}}{\Si;\epsilon}}, 
 {\Eblock{M_{1}\Orelse M_{2}}{\Si;\epsilon}\Orelse\Eblock{M_{3}}{\Si;\epsilon}})\big\} \\
&\cup&\big\{( {A\Orelse (B\Orelse C)}, {(A\Orelse B)\Orelse C}),\\
&& ( {A\Orelse \Eblock{M_{2}\Orelse M_{3}}{\Si;\epsilon}}, 
 {(A\Orelse\Eblock{M_{2}}{\Si;\epsilon})\Orelse\Eblock{M_{3}}{\Si;\epsilon}}), \\
&& ( {\Eblock{M_{1}}{\Si;\epsilon}\Orelse (\Eblock{M_{2}}{\Si;\epsilon}\Orelse C)}, 
 {\Eblock{M_{1}\Orelse M_{2}}{\Si;\epsilon}\Orelse C}), \\
&& ( {\Eblock{M_{1}}{\Si;\epsilon}\Orelse  (B\Orelse \Eblock{M_{3}}{\Si;\epsilon})}, 
 {(\Eblock{M_{1}}{\Si;\epsilon}\Orelse B)\Orelse\Eblock{M_{3}}{\Si;\epsilon}})\\
 &&\quad\big| \Eblock{M_{1}}{\Si;\epsilon}\To A,\,\Eblock{M_{2}}{\Si;\epsilon}\To B,\, \Eblock{M_{3}}{\Si;\epsilon}\To C\big\}\\
&\cup& \big\{( {\Eblock{M_{2}\Orelse M_{3}}{\Si';\epsilon}}, 
 {\Eblock{M_{2}}{\Si';\epsilon}\Orelse\Eblock{M_{3}}{\Si';\epsilon}}), ( {(D \Orelse E)},  {D\Orelse E})\\
 && \quad\big|  \Eblock{M_{1}}{\Si';\epsilon}\To \Eblock{\Retry}{\Si';\D},\, \Eblock{M_{2}}{\Si';\epsilon}\To D,\, \Eblock{M_{3}}{\Si';\epsilon}\To E\big\}\\
&\cup&\big\{ ( {F},  {F})\big| \Eblock{M_{1}}{\Si'';\epsilon}\To \Eblock{\Retry}{\Si';\D},\, \Eblock{M_{2}}{\Si'';\epsilon}\To \Eblock{\Retry}{\Si'';\D'},\, \Eblock{M_{3}}{\Si'';\epsilon}\To F\big\}~.\\
\end{array}\]

\item[\rulen{(absRt1)} $\alpha.\Retry  \AAbisim  \Retry$:]  suppose $\alpha=\Read a$:
\[
\begin{array}{rcl}
\mathcal R & = &\big\{( {\Eblock{\Read a.\Retry}{\Si;\epsilon}}, {\Eblock{\Retry}{\Si;\epsilon}})\big\}\\
&\cup&\big\{ ( {\Eblock{\Retry}{\Si;\Read a}},  {\Eblock{\Retry}{\Si;\epsilon}})\big| a^{n}\in \Si,\, n>0\big\}\\
&\cup&\big\{( {\Eblock{\Retry}{\Si;\epsilon}},  {\Eblock{\Retry}{\Si;\epsilon}})|  a\notin \Si\big\}~.\\
\end{array}\]

\item[\rulen{(absRt2)} $\Retry\Orelse M  \AAbisim  M \AAbisim M\Orelse \Retry$:] 
\[
\begin{array}{rcl}
\mathcal R_{1}&=&\big\{(\Eblock{\Retry\Orelse M}{\Si;\epsilon}, \Eblock{M}{\Si;\epsilon})\big\}\\
&\cup&\big\{(\Eblock{\Retry}{\Si;\epsilon}\Orelse A,A),\, (A,A)\big| \Eblock{M}{\Si;\epsilon}\To A\big\}\vspace{-0.5cm}\\
\\
\mathcal R_{2}&=&\big\{(\Eblock{M\Orelse \Retry}{\Si;\epsilon}, \Eblock{M}{\Si;\epsilon})\big\}\\
&\cup&\big\{ (A \Orelse\Eblock{\Retry}{\Si;\epsilon},A)\big| \Eblock{M}{\Si;\epsilon}\To A\big\}\\
&\cup&\big\{(\Eblock{\End}{\Si;\delta},\Eblock{\End}{\Si;\delta})\big| \Eblock{M}{\Si;\epsilon}\To \Eblock{\End}{\Si;\delta}\big\} \\
&\cup&\big\{(\Eblock{\Retry}{\Si;\epsilon},\Eblock{\Retry}{\Si;\delta})\big|\Eblock{M}{\Si;\epsilon}\To \Eblock{\Retry}{\Si;\delta}\big\}~.\\
\end{array}
\] 

\item[\rulen{(absEnd)} $\End\Orelse M \AAbisim  \End$:]  
\[
\begin{array}{rcl}
\mathcal R&=&\big\{(\Eblock{\End\Orelse M}{\Si;\epsilon},\Eblock{\End}{\Si;\epsilon}), (\Eblock{\End}{\Si;\epsilon},\Eblock{\End}{\Si;\epsilon})\big\}\\
&\cup& \big\{(\Eblock{\End}{\Si;\epsilon}\Orelse A,\Eblock{\End}{\Si;\epsilon})\;\big| \Eblock{M}{\Si;\epsilon}\To A\big\}~.\\
\end{array}
\]



\item[\rulen{(asy)} $\Inp a.\Out a  \Abisim \nil$:]  
\[
\mathcal R=\big\{(\Inp a. \Out a, \nil), \, (\Out a,\Out a),\,(\nil,\nil)\big\}~.
\]

\item[\rulen{(a-asy)} $\Atomic(\Read a.\Write a.\End)  \Abisim \nil$:]  
\[
\begin{array}{rcl}
\mathcal R&=&\big\{(\Atomic(\Read a.\Write a.\End), \nil), \, ( \Ablock{\Eblock{\Read a.\Write a.\End}{\Si;\epsilon}}{\Read a.\Write a.\End}, \nil)\big\}\\
&\cup&\big\{ ( \Ablock{\Eblock{\Write a.\End}{\Si;\Read a}}{\Read a.\Write a.\End}, \nil),\,( \Ablock{\Eblock{\End}{\Si;\Read a.\Write a}}{\Read a.\Write a.\End}, \nil)\big|a^{n}\in \Si, n>0  \big\}\\
&\cup& \big\{( \Ablock{\Eblock{\Retry}{\Si;\epsilon}}{\Read a.\Write a.\End}, \nil),\, (\Out a,\Out a),\,(\nil,\nil)\big| a\notin \Si\big\}~.\\
\end{array}\]

\item[\rulen{(a-1)} $\Atomic(\Read a.\End )  \Abisim \Inp a$:]  
\[
\begin{array}{rcl}
\mathcal R&=&\big\{(\Atomic(\Read a.\End), \Inp a), \, ( \Ablock{\Eblock{\Read a.\End}{\Si;\epsilon}}{\Read a.\End}, \Inp a)\big\}\\
&\cup&\big\{ ( \Ablock{\Eblock{\End}{\Si;\Read a}}{\Read a.\End}, \Inp a),\,(\nil,\nil)\big| a^{n}\in \Si, n>0\big\}\\
&\cup & \big\{( \Ablock{\Eblock{\Retry}{\Si;\epsilon}}{\Read a.\End}, \Inp a)\big| a\notin \Si\big\}~.\\
\end{array}\]
\end{description}


\section{Proof of Prposition~\ref{prop:normal-form}}\label{app:normal-form}
In this section we show that laws in Table~\ref{table:lawM} can be used for eliminating redundant branches from an atomic expression and obtaining an equivalent expression in normal form (see proof of Proposition~\ref{prop:normal-form}.) Some preliminary results are needed.

The next proposition states that if $K'$'s reads  include  $K$'s then $K'$ is bigger than $K$ in our weak atomic preorder.
\begin{proposition_a}\label{prop:retry}
Suppose $K=A_1.\cdots. A_n$ and $K'=B_1.\cdots .B_m$, with $A_i,B_j::=\Read a|\Write a$. If $\ReadL (K)\subseteq \ReadL (K')$ then $K\AAmore K'$.
\end{proposition_a}
\begin{proof}
It is enough to observe that if $\Eblock{K'}{\Si;\epsilon}\To\Eblock{\End}{\Si;\D}$ then $\ReadL (K')\subseteq \Si$ (rules \rulen{(ARdOk)} and \rulen{(ARdF)}); thus $\ReadL(K)\subseteq \Si$, and by \rulen{(ARdOk)} we get $\Eblock{K}{\Si;\epsilon}\To\Eblock{\End}{\Si;\D'}$.\qed
\end{proof}

As a consequence of the previous proposition, we obtain that, in an $\Orelse$ expression, a redundant branch, that is a branch which includes the reads of at least one of its preceding branches, can be eliminated.
\begin{proposition_a}\label{prop:delred}
Consider the expressions $K_1,\dots, K_n$ where, for $i=1,\dots,n$,  $K_i$ is  of the form $A_{i_1}.\cdots. A_{i_{n_i}}$ with $A_{i_j}::= \Read a|\Write a$. If $\ReadL (K_{j})\subseteq \ReadL (K_{n})$, for a $j$ such that $0<j<n$, then
\[
K_1 \Orelse \cdots \Orelse K_{n-1} \Orelse K_n
\AAbisim
K_1 \Orelse \cdots \Orelse K_{n-1}~.
\]   
\end{proposition_a}
\begin{proof}
The proof proceeds by using Proposition~\ref{prop:retry},  the fact that $M\sqcup M' \AAbisim M$ if and only if  $M\AAmore M'$ (see pag.~\pageref{AAmore}) and $\Orelse$'s rules in Table~\ref{operationalsemanticsmemory}.\qed
\end{proof}

As previously said, the proof of the following theorem show how to apply rules in Table~\ref{table:lawM} for rearranging an atomic expression into an equivalent one in normal form.
\begin{proposition_a}[Proposition~\ref{prop:normal-form}]
For every expression $M$ there is an expression $M'$ in normal form such that $M\AAbisim M'$.
\end{proposition_a}
\begin{proof}
The proof proceeds by induction on the structure of $M$:
\begin{description}
\item[$M=\End$:] $M'=M=\End$;

\item[$M=\Retry$:] $M'=M=\Retry$;

\item[$M=\alpha.N$:] by induction hypothesis, there is an $N'$ in normal form such that $N\AAbisim N'$. By Proposition~\ref{prop:Mcongpref}, $\alpha.N\AAbisim \alpha.N'$, thus by choosing $M'=\alpha.N'$ we obtain $M\AAbisim M'$;

\item[$M=N\Orelse N'$:] by induction hypothesis, there are  $N_0$ and $N'_0$, in normal form, such that $N\AAbisim N_0$ and $N'\AAbisim N'_0$. 
By Proposition~\ref{prop:Mcongorelse}, $M=N\Orelse N'\AAbisim N_0\Orelse N'_0$. We choose $M'$ by considering the following cases:
\begin{itemize}
\item if $N_0=\Retry$ we choose $M'=N'_0$, because, by \rulen{(absRt)}, $\Retry\Orelse N'_0\AAbisim N'_0$;


\item if $N_0= N_{0_1} \Orelse \dots\Orelse N_{0_n}$ and $N'_0= N'_{0_1} \Orelse \cdots\Orelse N'_{0_m}$, consider $P=\{j\,|\,k\in\{1,\dots, n\}:\;\ReadL (N_{0_k}) \subseteq \ReadL (N'_{0_j})\}$. If $P=\emptyset$ this means that $M'=N_0\Orelse N'_0$ is in normal form. 

Otherwise, suppose $P=\{j_1,\dots,j_l\}$ with $j_i<j_w$ for $i<w$; by applying Proposition~\ref{prop:delred} and~\ref{prop:Mcongorelse} and \rulen{(ass)} at every step, we have

\[
\begin{array}{l}
N_0 \Orelse N'_{0} \\
\AAbisim \quad(\mbox{by removing } N'_{0_{j_1}})\\
N_0 \Orelse N'_{0_1} \Orelse \cdots\Orelse N'_{0_{{j_1}-1}}\Orelse  N'_{0_{{j_1}+1}}\Orelse\cdots\Orelse N'_{0_m}\\
\AAbisim \quad(\mbox{by removing } N'_{0_{j_2}})\\
N_0 \Orelse N'_{0_1} \Orelse \cdots\Orelse N'_{0_{{j_1}-1}}\Orelse  N'_{0_{{j_1}+1}}\Orelse\cdots\\
\qquad\qquad\Orelse N'_{0_{{j_2}-1}}\Orelse  N'_{0_{{j_2}+1}}\Orelse\cdots \Orelse N'_{0_m}\\
\AAbisim \quad(\mbox{by removing } N'_{0_{j_3}})\\
\vdots\\
\AAbisim \quad(\mbox{by removing } N'_{0_{j_l}})\\
N_0 \Orelse N'_{0_1} \Orelse \cdots\Orelse N'_{0_{{j_1}-1}}\Orelse  N'_{0_{{j_1}+1}}\Orelse\cdots\\
\qquad\qquad\Orelse N'_{0_{{j_2}-1}}\Orelse  N'_{0_{{j_2}+1}}\Orelse\cdots \Orelse N'_{0_{{j_l}-1}}\\
\qquad\qquad\Orelse  N'_{0_{{j_l}+1}}\Orelse\cdots\Orelse N'_{0_m}\\
=\; M' \mbox{ (that is in normal form.)}
\end{array}
\]

\end{itemize}
In every case, $M'\AAbisim N_0\Orelse N'_0$, thus $M\AAbisim M'$.\qed
\end{description}
\end{proof}


\section{Proofs of Section~\ref{sec:trace}}\label{app:may}

\begin{lemma_a}[Lemma~\ref{lemma:toreduction}]\label{lemmaapp1}
Assume that $s' \tot s$ and $P\Tolabel{\Out s} P'$, then there is a process $P''$ such that $P\Tolabel{\Out {s}'} P''$.
\end{lemma_a}
\begin{proof}
$s'\tot s$ means $s' \tob^{n}s$, for some $n\geq 0$. The proof proceeds by induction on $n$. For $n=0$ we have $s=s'$. Suppose $n>0$ and $s' \tob^{{n-1}}s'' \tob s$. The result follows 
by induction hypothesis if we show that $P\Tolabel{\overline{s''}}$. We proceed by distinguishing the possible cases for $s''\tob s$ according to laws \rulen{(TO1)}-\rulen{(TO4)}.
\begin{description}
\item[\rulen{(TO1)}] $s''=rr' $ and $s=r \{a\} r'$, thus $\overline {s''}=\overline r \overline{r'}$ and $\overline s=\overline r \overline a \overline {r'}$. 
$P\Tolabel {\overline s}$ implies $P\Tolabel{\overline r}P_{1} \Tolabel{\overline a}P_{2}\Tolabel{\overline {r'}}$, and by Proposition~\ref{prop:out}, 
$P_{1}\sim P_{2}\,|\, \Out a$, that is $P\Tolabel{\overline r}P_{2}\,|\,\Out a \Tolabel{\overline{r'}}$, hence $P\Tolabel{\overline{s''}}$;

\item[\rulen{(TO2)}] $s''=rl\{a\}r' $ and $s=r \{a\}l r'$, thus $\overline {s''}=\overline {r}\overline l \overline a\overline{r'}$ and $\overline s=\overline r \overline a\overline l \overline {r'}$. 
$P\Tolabel {\overline s}$ implies $P\Tolabel{\overline r}P_{1} \Tolabel{\overline a}P_{2}\Tolabel{\overline l}P_{3}\Tolabel{\overline {r'}}$, and by Proposition~\ref{prop:out}, 
$P_{1}\sim P_{2}\,|\, \Out a$, that is $P\Tolabel{\overline r}P_{2}\,|\,\Out a \Tolabel{\overline{l}}P_{3}\,|\,\Out a \Tolabel{\overline a}P_{3}\Tolabel{\overline {r'}}$, hence $P\Tolabel{\overline{s''}}$;

\item[\rulen{(TO3)}] $s''=r r'$ and $s=r \{a\} \overline a r'$, thus $\overline{s''}=\overline r\overline{r'}$ and $\overline s =\overline r \overline a \{a\} \overline{r'}$.
$P\Tolabel {\overline s}$ implies $P\Tolabel{\overline r}P_{1}\Tolabel{\overline a}P_{2}\Tolabel a P_{3}\Tolabel {\overline {r'}}$, hence, by Proposition~\ref{prop:out}, $P_{1}\sim P_{2}\,|\, \Out a$, 
that is $P_{2}$ can synchronize with $\Out a$ and $P\Tolabel{\overline r}P_{2}\,|\,\Out a\To P_{3}\Tolabel {\overline {r'}}$, that is $P\Tolabel{\overline{s''}}$;

\item[\rulen{(TO4)}] $s''=\{a_{1}\}\cdots \{a_{n}\}$ and $s=\{a_{1},\cdots,a_{n}\}$, or viceversa; in this case $\overline{s}=\overline{s''}$ by definition of $\overline\cdot$.\qed
\end{description} 
\end{proof}

\begin{lemma_a}[Lemma~\ref{lemma:observer}]\label{lemmaapp2}
Consider two traces $s$ and $r$. If there is a process $Q$ such that $\observer(s) \Tolabel{\Out r} \, \Tolabel{\Out w} Q$ then $r\tot s$.
\end{lemma_a}
\begin{proof}
The proof proceeds by induction on $s$.
\begin{description}
\item[$s=\overline a s'$:] $\observer(s)=\Inp a.\observer(s')$ and $\observer(s)\Tolabel{\overline r\overline w}$ implies $\overline r=a\overline r'$ such that 
$\observer(s)\tolabel{\{a\}}\observer(s')\Tolabel{\overline{r'}}$. By induction hypothesis, $r'\tot s'$, hence by prefixing, $r=\overline a r' \tot \overline a s'=s$;

\item[$s=\{a_{1},\cdots,a_{n}\}s'$:] $\observer(s)=\big(\prod_{a\in\{a_{1},\cdots,a_{n}\}}\Out a\big)\,|\, \observer(s')$. We have $\observer(s)\Tolabel{\overline r\overline w}$, we can distinguish 
the following cases depending on $\overline r$:
\begin{description}
\item[$\overline {a_{i}}\notin \overline r$:] by induction hypothesis, $\observer(s')\Tolabel{\overline r \overline w}$ implies $r\tot s'$ and by \rulen{(TO1)}, 
$r\tot s' \tot \{a_{1}\}\cdots \{a_{n}\}s' \:{}_0\succeq\tob\{a_{1},\cdots,a_{n}\}s'=s$;

\item[$\overline {a_{i_{1}}},\cdots \overline {a_{i_{k}}}\in \overline r$ for $\{a_{i_{1}},\cdots, a_{i_{k}}\}\subseteq \{a_{1},\cdots,a_{n}\}$:] in this case 
$\overline r=\overline{r_{1}}\overline {a_{i_{1}}}\cdots \overline{r_{k}}\overline {a_{i_{k}}}\overline{r_{k+1}}$ and 
$\observer(s')\Tolabel{\overline{r_{1}} \cdots \overline{r_{k+1}}\overline{w}}$. By induction hypothesis, $r_{1}\cdots r_{k+1}\tot s'$:
\[
\begin{array}{rcll}
r&=& {r_{1}} \{a_{i_{1}}\}\cdots{r_{k}} \{a_{i_{k}}\}{r_{k+1}}\\
&\tot& \{a_{i_{1}}\}\cdots \{a_{i_{k}}\}r_{1}\cdots r_{k+1} & (\textrm{by \rulen{(TO2)}})\\
& \tot& \{a_{i_{1}}\}\cdots \{a_{i_{k}}\} s' & (\textrm{by induction and prefixing})\\
& \tot& \{a_{1}\}\cdots \{a_{n}\}s'  & (\textrm{by \rulen{(TO1)} and \rulen{(TO2)}})\\
& {}_0\!\succeq\tob & \{a_{1}\cdots a_{n}\}s'&(\textrm{by \rulen{(TO4)}})\\
&=&s;
\end{array}
\]

\item[$\overline r=\overline{r_{1}}\cdots \overline{r_{k}}$ and $\observer(s')\Tolabel{\overline{r_{1}}\{a_{i_{1}}\}\cdots \overline{r_{k}}\{a_{i_{k}}\}r_{k+1}}$ for $\{a_{i_{1}},\cdots,a_{i_{k}}\}\subseteq \{a_{1},\cdots ,a_{n}\}$:]
by induction hypothesis, ${r_{1}}\overline{ a_{i_{1}}}\cdots {r_{k}} \overline {a_{i_{k}}} r_{k+1}\tot s'$ and:
\[
\begin{array}{rcll}
r&=&r_{1}\cdots r_{k}\\
&\tot &r_{1}\{a_{i_{1}}\}\overline{a_{i_{1}}} \cdots r_{k} \{a_{i_{k}}\}\overline{a_{i_{k}}} r_{k+1} & (\textrm{by \rulen{(TO3)}})\\
&\tot& 
\{a_{i_{1}}\}\cdots \{a_{i_{k}}\} {r_{1}}\overline{ a_{i_{1}}}\cdots {r_{k}} \overline {a_{i_{k}}}  r_{k+1} & (\textrm{by \rulen{(TO2)}})\\
&\tot& 
\{a_{i_{1}}\}\cdots \{a_{i_{k}}\} s'& (\textrm{by induction})\\
&\tot& \{a_{1}\}\cdots \{a_{n}\}s_{0}& (\textrm{by \rulen{(TO1)} and \rulen{(TO2)}})\\
&{}_0\!\succeq\tob&\{a_{1}\cdots a_{n}\}s' & (\textrm{by \rulen{(TO4)}})\\
&=&s.
\end{array}
\]\qed
\end{description}
\end{description}
\end{proof}

The proof of the full-abstraction theorem is standard (see e.g.~\cite{BDNP}).

\begin{theorem_a}[Theorem~\ref{th:coincidence}]
For all processes $P$ and $Q$, $P\maypr Q$ if and only if $P\Ll Q$.
\end{theorem_a}
\begin{proof}
\begin{description}
\item[$\Rightarrow$:] Suppose $P\Ll Q$ and $P\; \mathit{may}\; O$ for any observer $O$ we have to show that $Q\;\mathit{may}\; O$. $P\;\mathit{may}\; O$ 
means that $P\,|\, O\Tolabel{\overline w}$, that is there exists a trace $s$ such that $P\Tolabel s$ and $O\Tolabel{\overline s\overline w}$. 
$P\Ll Q$ implies that there exists $s'\tot s$ such that $Q\Tolabel{s'}$. $s'\tot s$ implies $s'w \tot s w$. By Lemma~\ref{lemmaapp1} and $O\Tolabel{\overline s \overline w}$ we 
get that $O\Tolabel{\overline{s'}\overline w}$. Hence, from $Q\Tolabel{s'}$ we obtain $Q\,|\,O\Tolabel{\overline w}$, that is $Q\;\mathit{may} \;O$ ($P\maypr Q$).

\item[$\Leftarrow$:] Suppose $P\maypr Q$ and $P\Tolabel{s}$, we have to show that there exists $s'\tot s $ such that $Q\Tolabel{s'}$. From $P\Tolabel{s}$ and 
$\observer(s)\Tolabel{\overline s\overline w}$ we have $P\,|\,\observer(s)\Tolabel{\overline w}$, that is $P\;\mathit{may}\; \observer (s)$. Hence 
$Q\;\mathit{may}\;\observer (s)$, that is $Q\,|\,\observer (s)\Tolabel{\overline{w}}$. Thus, there exists $s'$ such that $Q\Tolabel{s'}$ and $\observer(s)\Tolabel{\overline{s'}\overline w} $, and, by Lemma~\ref{lemmaapp2} and $\observer(s)\Tolabel{\overline{s'}\overline w}$ we have $s'\tot s$, that is $P\Ll Q$.\qed
\end{description}
\end{proof}

\begin{lemma_a}[Lemma~\ref{lemma:nfend}]\label{lemmaapp3}
Assume $M = \bigsqcup_{i \in 1..n}K_{i}$ is an expression in normal form. For every index $i$ in $\{1,\dots,n\}$ we have $\Atomic(M) \pv \Si_{i} \to^* \Ablock{\Eblock{\End}{\sigma_{i} ; \delta}}M \pv  \sigma_i$ where $\sigma_{i} = \ReadL {(K_{i})} = \ReadL(\delta)$ and $\WriteL(\delta) = \WriteL(K_{i})$.
\end{lemma_a}
\begin{proof}
By definition of normal form.\qed
\end{proof}

\begin{corollary_a}\label{corollary:nfend}
Assume $M = \bigsqcup_{i \in 1..n}K_{i}$ is an expression in normal form. The possible behavior of $\Atomic(M)$ can be described as 
$\Atomic(M)\Tolabel{{\Si_{i}}}\prod_{b\in \WriteL(r_{i})}b$ for every $i\in 1..n$ where $\Si_{i}$ is the multiset $\ReadL(K_{i})$.
\end{corollary_a}
\begin{proof}
By Lemma~\ref{lemmaapp3}, rule \rulen{(atOk)} and definition of $\tolabel\mu$.\qed
\end{proof}

We can prove now the main result of Section~\ref{sec:trace}, that is that may-testing semantics is not able to distinguish the behavior of an atomic expression from the behavior of the corresponding CCS process.

\begin{theorem_a}[Theorem~\ref{th:may}]
For every expression $M$ in normal form we have $\Atomic(M) \mayeq\tradm M$.
\end{theorem_a}
\begin{proof}
The proof proceeds by using the alternative preorder instead of the may preorder; in what follows it is shown that:
\begin{enumerate}
\item $\Atomic(M) \Ll \tradm M$;
\item $\tradm M\Ll \Atomic(M)$.
\end{enumerate}
Remember that $M$ is in normal-form, thus $M=\mathrm{OrElse}_{i=1,\dots, n}K_{i}$ and $\tradm M=\sum_{i=1,\dots,n}\tradm {K_{i}}$.
The two points are shown in what follows. 
\begin{enumerate}
\item 
For proving that $\Atomic(M) \Ll \tradm M$, we have to show that $\forall s $ such that $\Atomic(M)\Tolabel s$ there exists $s'\tot s$ such that 
$\tradm M\Tolabel{s'}$. We distinguish the following cases for $s$:
\begin{description}
\item[$s=\epsilon$:] in this case we can choose $s'=\epsilon$;

\item[$s=\delta \overline a_{i_{1}}\cdots\overline a_{i_{l}}$ with $l\geq 0$:] by Corollary~\ref{corollary:nfend}, there is a $j\in\{1,\dots, n\}$ such that 
$\delta=\ReadL(K_{j})$, 
\[
\Atomic(M)\Tolabel{{\ReadL (K_{j})}}\Out{a_{1}}\,|\,\cdots\,|\,\Out{a_{m}}\Tolabel{\overline a_{i_{1}}\cdots\overline a_{i_{l}}}
\]
with $\{a_{i_{1}},\cdots,a_{i_{l}}\}\subseteq\{a_{1},\cdots,a_{m}\}=\WriteL(K_{j})$. 

Suppose $\ReadL (K_{j})=\{b_{1},\cdots,b_{k}\}$. By definition, 
$\tradm {K_{j}}=  \Inp{b_{1}}.\cdots.\Inp{b_{k}}.(\Out{a_{1}}\,|\, \cdots\,|\,\Out{a_{m}})$ with 
$\{a_{1},\cdots,a_{m}\}=\WriteL(K_{j})$.  
That is, if we choose the $j$-th summands of $\tradm M$, we have $\tradm M\Tolabel{s'}$ with $s'=\{b_{1}\}\cdots \{b_{k}\}\overline {a_{i_{1}}}
\cdots \overline{a_{i_{l}}}$, and by \rulen{(TO4)} $s'{}_0\!\succeq\tob s$;  
\end{description} 

\item
For proving that $\tradm M\Ll \Atomic(M)$, we have to show that $\forall s $ such that $\tradm M\Tolabel s$ there exists $s'\tot s$  such that 
$\Atomic(M)\Tolabel{s'}$. We distinguish the following cases for $s$:
\begin{description}
\item[$s=\{b_{1}\}\cdots \{b_{k}\}$:] $s $ contains only input actions, thus we can choose $s'=\epsilon \tot s$ and $\Atomic(M)\Tolabel{s'}$;

\item[$s=\{b_{1}\}\cdots \{b_{k}\}\overline{a_{1}}\cdots\overline {a_{m}}$ with $m>0$:] in this case there is a $j\in\{1,\dots,n\}$ such that 
$\tradm{K_{j}}\Tolabel s$, $\{b_{1},\cdots, b_{k}\}=\ReadL (K_{j})$ 
and $\{a_{1},\cdots,a_{m}\}\subseteq \WriteL(K_{j})$ 
(by definition of $\tradm\cdot$). 
Suppose $\sigma=\ReadL (K_{j})$, by Lemma~\ref{lemmaapp3}, 
$\Atomic(M);\sigma \To \Ablock{\Eblock{\End}{\sigma;\delta}}M$ with  $\ReadL(\delta)=\ReadL (K_{j})$ and 
$\WriteL(\delta)=\WriteL(K_{j})$. This means that 
$\Atomic(M)\Tolabel{{\ReadL (K_{j})}} \prod_{a\in \WriteL(K_{j})}\Out  {a}$, that is (by  \rulen{(TO4)}) there is an 
$s'=\ReadL (\delta) \overline{a_{1}}\cdots \overline{a_{m}}\;{}_0\!\succeq\tob\{ b_{1}\}\cdots \{b_{k}\}\overline{a_{1}}\cdots \overline{a_{m}}= s$ such that $\Atomic(M)\Tolabel{s'}$.\qed
\end{description}

\end{enumerate}
\end{proof}

\end{appendix}

\end{document}